# Buzz Buzz: Haptic Cuing of Road Conditions in Autonomous Cars for Drivers Engaged in Secondary Tasks

By

Shivam Pandey

# RICE UNIVERSITY

Buzz Buzz: Haptic Cuing of Road Conditions in

Autonomous Cars for Drivers Engaged in Secondary Tasks

By

# Shivam Pandey

A THESIS SUBMITTED

IN PARTIAL FULFILLMENT OF THE

REQUIREMENT FOR THE DEGREE

# Doctor of Philosophy

APPROVED, THESIS COMMITTEE:

| | |
|---|---|
| Michael Byrne | Patricia DeLucia |
| Professor, Psychological Sciences and Computer Science | Professor, Psychological Sciences |
| Philip Kortum | Marcia O'Malley |
| Associate Professor, Psychological Sciences | Professor, Mechanical Engineering |

Houston, Texas

April 2021

**Abstract**

Can drivers' situation awareness during automated driving be maintained using haptic cues that provide information about road and traffic scenarios while the drivers are engaged in a secondary task? And can this be done without disengaging them from the secondary task? Multiple Resource Theory predicts that using different sensory channels can improve multiple-task performance. Using haptics to provide information avoids the audio-visual channels likely occupied by the secondary task. An experiment was conducted to assess whether drivers' situation awareness could be maintained using haptic cues. Drivers played Fruit Ninja as the secondary task while seated in a driving simulator with a Level 4 autonomous system driving. A mixed design was used for the experiment with the presence of haptic cues and the presentation time of situation awareness questions as the between-subjects conditions. Five road and traffic scenarios comprised the within-subjects part of the design. Subjects who received haptic cues had a higher number of correct responses to the situation awareness questions and looked up at the simulator screen fewer times than those who were not provided cues. Subjects did not find the cues to be disruptive and gave good satisfaction scores to the haptic device. Additionally, subjects across all conditions seemed to have performed equally well in playing Fruit Ninja. It appears that haptic cuing can maintain drivers' situation awareness during automated driving while drivers are engaged in a secondary task. Practical implications of these findings for implementing haptic cues in autonomous vehicles are also discussed.

## ACKNOWLEDGEMENTS

I would like to begin by thanking Hanna Gratch and Luna Cortelezzi for their help with data collection for my dissertation during the COVID-19 pandemic. In-person data collection during the pandemic required extensive precautions and contact tracing. Without their help, I would not have been able to collect the data for my dissertation. I am also grateful to my advisor, Dr. Mike Byrne, for his guidance and his willingness to allow me to experiment with various areas of human factors research during my doctoral studies. I wish to thank Dr. Pat DeLucia, Dr. Phil Kortum, and Dr. Marcie O'Malley who provided me guidance and input, as members of my dissertation committee, to help refine my dissertation idea into its final form. Additionally, I am also grateful to Dr. Pat DeLucia for giving me access to her driving simulator to record the drive videos. I would like to thank students in the MAHI Lab for providing me the software and hardware to design and display the haptic cues. I would like to acknowledge the funding provided by Rice Social Sciences Research Institute for my dissertation and pre-dissertation research. I am also thankful for all the love and support my friends and family have provided me.

TABLE OF CONTENTS

LIST OF TABLES

# LIST OF FIGURES

# ACRONYMS

Attack Sustain Release: ASR
False Discovery Rate: FDR
Engineering Acoustics Inc.: EA
General Motors: GM
Human-Machine Interface: HMI
Non-Driving Related Task: NDRT
Personal Protective Equipment: PPE
Situation Awareness Global Assessment Technique: SAGAT
Society of Automotive Engineers: SAE
System Usability Scale: SUS
Takeover request: TOR

## Chapter 1
## Introduction

Can drivers' situation awareness during automated driving be maintained using haptic cues that provide information about road and traffic scenarios while the drivers are engaged in a secondary task? And can this be done without disengaging them from the secondary task? Driver assistance and advanced safety systems are becoming an increasingly common offering among automobile manufacturers (Daily et al., 2017; Endsley, 2019). Autonomous driving systems based on these driver assistance and safety systems have been widely reported in the news. Auto manufacturers have branded these autonomous systems with catchy names such as Tesla's Autopilot and General Motors' (GM) Super Cruise. Manufacturers have integrated driver assistance and advanced safety systems together to different extents to offer systems with varying levels of autonomous driving capability (Daily et al., 2017; Endsley, 2019). Beyond the auto manufacturers, there are several companies developing autonomous driving systems and technologies related to autonomous driving systems such as Waymo and Aptiv with their self-driving taxis, and Intel's efforts in deep-learning for collision avoidance by acquiring Mobileye (Daily et al., 2017).

**Secondary Tasks**

Autonomous driving systems promise to free the driver from the task of actively driving the car to varying degrees depending on system capabilities. This creates problems with loss of situation awareness in the driver since the driver is no longer actively involved in driving and may instead be engaged in secondary tasks such as tasks such as reading, texting, watching videos, playing games, or sleeping (Cunningham & Regan, 2018; Endsley, 2019; Telpaz et al., 2017; Wan & Wu, 2018a, 2018b). In a literature review covering driver inattention in autonomous vehicles, authors identified several reasons as to why drivers tend to take up secondary tasks during automated driving including the nature of the driving task changing to a passive supervisory task, monotony, and boredom (Cunningham & Regan, 2018). The literature has also provided evidence that the tendency to take up secondary tasks during automated driving exists both in a driving simulator and in the real world (Cunningham & Regan, 2018). Researchers have also found the previously mentioned secondary tasks common among passengers using public transportation, and also common in a large opinion survey covering secondary tasks in fully autonomous vehicles (Wan & Wu, 2018a, 2018b). The human tendency to take up secondary tasks when automation is engaged is not limited to driving only, human-automation interaction research in aviation has also uncovered similar findings (Telpaz et al., 2017).

**Levels of Automation**

Due to differences in the autonomous capabilities of cars offered by various manufacturers, it is critical to understand the different levels of automation and what they mean before a sound discussion can be conducted on the implications of such autonomous systems. The Society of Automotive Engineers (SAE) defines six levels of automation, which span from full human control with no automation to full autonomous control (SAE International, 2018). Some examples of Level 0 systems include automatic emergency braking, blind spot warning, and lane departure warning. Level 1 systems include either lane centering or adaptive cruise control (SAE International, 2018). Tesla's Autopilot is an example of a Level 2 system since it has lane centering and adaptive cruise control at the same time (Endsley, 2019; SAE International, 2018). Higher level systems such as Level 3 traffic jam navigators, and Level 4 systems such as driverless taxis are still under development (Endsley, 2019; SAE International, 2018). A Level 5 system would be capable of driving anywhere in any condition (SAE International, 2018). There are several key takeaways from these definitions: first, Level 0-2 automation technology is primarily defined in terms of driver assistance, with the driver largely expected to perform driving functions. Second, at Level 5 the driver's ability to respond appropriately is not critical to system performance. Third, more commonly at Level 3 and depending on the specific capabilities of the autonomous system implemented, even at Level 4, the drivers may be required to take over control.

## Situation Awareness

Since the loss of situation awareness can occur in drivers during periods of automated driving as they take up secondary tasks, it is important to define situation awareness prior to discussing the loss of situation awareness during automated driving and possible countermeasures to address the loss of situation awareness. Situation awareness is defined as (Endsley, 1995a):

> The perception of the elements in the environment within a volume of time and space, the comprehension of their meaning, and the projection of their status in the near future.

Decreased situation awareness and decreased performance in drivers can occur while automation is engaged due to several reasons including poor vigilance and over-reliance, limited information about the autonomous system or the environment, and low cognitive engagement since drives are no longer actively involved in driving the vehicle (Endsley, 2019). These issues become more severe in the presence of Non-Driving Related Tasks (NDRT) (Endsley, 2019). Prior research efforts in the use of Human-Machine Interface (HMI) in autonomous cars to address challenges due to the loss of situation awareness in drivers have focused on multiple levels of automation, on unimodal and multimodal TORs, on what happens once a TOR has been issued, and on using haptic cues for warning/informational and guidance purposes.

**Takeover Requests**

Prior to designing any new solutions that can maintain driver situation awareness during automated driving with the driver engaged in NDRTs, it is important to first understand how prior research in HMI has addressed challenges with transfer of control to manual driving and what strategies have been used in prior research to alert drivers that an intervention is necessary. Autonomous driving systems can free the driver from the task of actively driving the car to varying degrees depending on system capabilities allowing the driver to take up secondary tasks not related to driving. However, even with the automation engaged, drivers may be asked to take over control of the vehicle from the autonomous driving system in case of system malfunctions, such as hardware or software errors, or if the autonomous system encounters a situation outside of system limits that it may not be able to handle, such as a deer on the road or degraded driving conditions due to inclement weather. This is known as a takeover request (TOR) (Gold et al., 2016; Naujoks et al., 2018, 2019; Yun & Yang, 2020). TORs may be implemented as unimodal or multimodal requests. A unimodal TOR uses only one of the sensory channels and may be displayed as a visual warning, an audio alert, or a vibrotactile alert (Gold et al., 2016). Multimodal TORs incorporate a combination of one or more sensory channels such as an audio-visual, audio-vibrotactile, or an audio-visual-vibrotactile TOR (Eriksson & Stanton, 2017a, 2017b; Naujoks et al., 2015; Politis et al., 2015; Yun & Yang, 2020).

I have provided an overview of what a TOR is, when it may be issued, and what forms a TOR may take. The next step is to discuss some examples of how TORs have been implemented in prior research and what kind of experiment conditions have been investigated. This discussion will help gain an understanding of how to design a new solution to maintain driver situation awareness that would work well with TORs in an autonomous vehicle but without being confused for a TOR. In a study that used unimodal audio TORs investigating the effect of traffic density and verbal tasks on takeover performance in highly automated driving, researchers found that the presence of traffic lowered the takeover quality and increased takeover times (Gold et al., 2016). The verbal task did not have an effect on takeover time but had minor effects on takeover quality (Gold et al., 2016).

Prior research has also included multimodal TORs, for instance, researchers have used audio-visual TORs created by combining an image displayed in the instrument panel and an audio request to resume control in studies investigating transition to manual control from automated driving (Eriksson & Stanton, 2017a, 2017b). In another study, researchers implemented multimodal TORs as a message displayed in the instrument panel and an audio warning when researching controllability of system limits during automated driving (Naujoks et al., 2015). Multimodal TORs combining one or more sensory channels including haptics have also been implemented in prior research. In a study investigating unimodal and multimodal TORs using all combinations of audio, visual, and vibrotactile TORs, researchers found multimodal TORs of high urgency had quicker transitions to manual control and lower quality of takeover with unimodal visual warnings (Politis et al., 2015). Researchers reported similar findings in another study comparing unimodal and multimodal TORs created using one or more of the audio, visual, and vibrotactile warnings, and found that the tri-modal warning showed the best performance overall whereas the visual TOR had the worst performance (Yun & Yang, 2020). Thus, prior research indicates that TORs are likely to be multimodal, therefore designing a unimodal solution to maintaining driver situation awareness during automated driving would make it less likely for drivers to misidentify it as a TOR.

**Haptic Cuing**

Continuing with the discussion of how HMI has been used in previous research, it also important to define haptic cues and discuss how researchers have used haptic cuing in vehicles for warning, informational, and guidance purposes, prior to designing a new solution that can maintain driver situation awareness during automated driving with the driver engaged in NDRTs. Haptic cues are vibration stimuli that rely on our sense of touch to provide information. Haptic cues are picked up by tactile receptors in the skin, called mechanoreceptors, which sense pressure, vibration, and slip (K. E. MacLean, 2008). Haptic cues can be programmed to have a variety of frequencies, amplitudes, and durations to generate a wide range of vibrotactile stimuli (K. E. MacLean, 2008; K. MacLean & Enriquez, 2003). Haptic feedback has been used in prior research to assist drivers with warning, informational, and guidance purposes. In research investigating informational haptic feedback, previous efforts have created cues associated with multiple priority levels and tested them in varying task conditions in an effort to support drivers' attention management (Cao et al., 2010). Drivers learned the cues quickly and were able to identify them quickly and accurately (Cao et al., 2010). In another study, researchers provided continuous informational haptic feedback on the gas pedal about inter-vehicle separation to drivers during car following and found that tasks such as selecting tracks on a CD player were performed more quickly in more critical car following scenarios without compromising safety by maintaining the same separation from the lead vehicle as drivers without haptic support (Mulder et al., 2010).

Research investigating haptic cues for guidance purposes has involved providing haptic feedback on the steering wheel to assist drivers in lane keeping during curve negotiation (de Nijs et al., 2014). The same study found that haptic feedback was best at addressing the loss of near visual information (de Nijs et al., 2014). Studies have also investigated detrimental effects of haptic feedback on the steering wheel which may lead speeding behavior in drivers and discussed modifications to the feedback that help address those issues while realizing the safety benefits of haptic feedback (Melman et al., 2017).

Prior research investigating haptic cues for warning purposes has included providing multimodal warnings to drivers to convey varying levels of urgency. Warning using a higher number of modalities were perceived as more urgent but also as more annoying; however annoyance had a lower effect compared to urgency (Politis et al., 2013). Researchers have also used Speech Tactons, which are vibrotactile equivalents of spoken messages, to provide warning of varying levels of urgency to drivers and found that drivers recognized the tactons when displayed separately or when displayed as a part of multimodal warnings (Politis et al., 2014). Research has also been conducted on using haptic cues in TORs to improve takeover performance and found haptic cuing to be beneficial to the transfer of control (Politis et al., 2015; Telpaz et al., 2017; Wan & Wu, 2018b; Yun & Yang, 2020). Therefore, prior research in using haptic cuing as an HMI modality for vehicles has consistently shown promising results for drivers in a wide range of HMI implementations which included providing warnings, information, and guidance to drivers. Next, it is worth discussing driver interaction with autonomous systems to get a general understanding of how drivers behave in autonomous vehicles prior to designing any solution to help maintain their situation awareness during automated driving.

**Driver Interaction with Autonomous Systems**

After a discussion of TORs and haptic cuing as they relate to HMI, the next critical step is to understand driver interaction with autonomous systems to ensure any solution designed to maintain driver situation awareness during automated driving can do so effectively and without having a negative impact on the drivers' experience. As the level of autonomy increases, drivers are more likely to be distracted as they take up secondary tasks since a vehicle with higher autonomy requires less driver input (de Winter et al., 2014; Merat et al., 2014; Merat & Jamson, 2009). In a meta-analysis and narrative review of the literature, authors found that drivers in automated vehicles had about half the self-reported workload compared to manual driving on the NASA Task Load Index or Rating Scale Mental Effort (de Winter et al., 2014). In the same analysis researchers found that based on 12 studies, drivers in autonomous vehicles completed more than 2.5 times the number of tasks on an in-vehicle display than the drivers who were driving manually (de Winter et al., 2014). In another study comparing the effects of highly automated driving on driver behavior, the researchers found that drivers with automation engaged responded to critical events much later than those driving manually even with an audio TOR (Merat & Jamson, 2009). The authors suggested that the delay in response may be explained by the lower situation awareness of the drivers in the highly automated experiment condition, and perhaps also because they may have waited for the audio TOR before responding to critical events (Merat & Jamson, 2009).

In a chapter summarizing a series of studies investigating driver behavior and secondary task uptake in automated vehicles, researchers found that takeover performance suffered when drivers were engaged in NDRTs compared to when they paid attention to the road (Merat et al., 2014). In the same series of studies, the researchers found evidence that drivers engaged in a higher number of NDRTs such as watching a DVD, reading, magazines, and eating (Merat et al., 2014). Researchers also investigated drivers' visual attention in the same series of studies and found their visual attention became more dispersed with increased automation. Drivers paid less attention to the road center during automated driving and their fatigue levels also rose with the increase in automation levels (Merat et al., 2014).

It has been observed that drivers using a Level 2 autonomous system may sometimes take up secondary tasks, thus behaving as if they were using a Level 3 system (Dogan et al., 2017). The researchers discuss further that the change in the nature of the task from active control to passive monitoring may be one of the primary reasons for the decrease in situation awareness. In fact, they note that once drivers are removed from active control, their situation awareness may fall below system designer expectations (Dogan et al., 2017). The researchers also identify drivers' tendency to assume that the autonomous system would not fail to be one of the reasons for decreased monitoring by the drivers (Dogan et al., 2017). Another study similarly found that drivers overestimate the reliability of autonomous systems thereby placing higher than warranted levels of trust in the system (Banks et al., 2018). In the same study, the researchers investigated whether a Level 2 system could support the drivers' monitoring role since they were no longer actively driving. They found that the autonomous system could not support drivers' monitoring function properly and drivers became complacent and over trusted the autonomous system (Banks et al., 2018). Studies have shown that issues such as these create problematic circumstances where drivers may suffer from the loss of situation awareness as they become increasingly disengaged from the task of driving the vehicle, and hence may not be able to provide an appropriate and timely response to a TOR by the vehicle (Cunningham & Regan, 2018). Researchers have also discussed the concept of automation conundrum, which implies that (Endsley, 2019):

> The more automation is added to a system, and the more reliable and robust the automation is, the less likely that human operators overseeing the automation will be aware of critical information and able to take over manual control when needed.

Numerous studies have shown that as automation capabilities increase, users are more likely to lose situation awareness and place unwarranted levels of trust in the system thereby decreasing their ability to oversee and intervene effectively (Endsley, 2019). Prior research has studied various aspects driver performance in transfer of control once a TOR has been issued. One of the studies determined that self-paced transition to manual control resulted in better driver takeover performance compared to system-paced transfer of control (Eriksson & Stanton, 2017b). This finding is particularly relevant to higher levels of automation, such as Level 4, where the system could alert the driver well in advance of a system boundary to allow the driver to complete a self-paced transition to manual control (Eriksson & Stanton, 2017b). Research has also found that drivers engaged in secondary tasks respond slower to TORs in non-critical scenarios and show a larger variance in their response—transitions to manual control need to account for this larger variance among drivers engaged in secondary tasks and system designers should not limit themselves to considerations of mean and median response times only (Eriksson & Stanton, 2017a).

Thus far I have discussed TORs and haptic cuing as they relate to HMI, and issues surrounding driver interaction with autonomous systems. They have each provided a part of the picture, however, prior to designing any solution to maintain driver situation awareness during automated driving another important piece of the puzzle needs to be discussed, that is the part covering how NDRTs may engage drivers during automated driving. Since drivers in autonomous cars are likely to take up NDRTs such as reading, texting, watching videos, playing games, or sleeping, it is important to understand how NDRTs may engage the drivers before considering how to best maintain their situation awareness. It is important to understand how NDRTs may engage drivers before creating a solution to maintain their situation awareness, otherwise there is a risk of developing an ineffective solution because drivers may not notice the solution or because the solution may interfere with the NDRT. Multiple resource theory can help with understanding how NDRTs engage drivers and help identify ways to best maintain driver situation awareness while attempting to minimize interference with NDRTs.

## Multiple Resource Theory

Multiple resource theory is a guide to understanding and predicting time-sharing, or multi-task, performance in how people perceive and respond to stimuli (Wickens, 2002, 2008). The word "multiple" implies parallel or fairly independent processing ability, and the idea of resources implies that factors that drive multi-task performance are both limited and allocatable (Wickens, 2002).

Multiple resource theory is based on the 4-D multiple resource model. The 4-D model further defines the resource components of the multiple resource theory, which have been supported by neurophysiological evidence (Wickens, 2008). The first dimension is stages of processing that includes perception, cognition, and response, indicating that perception and cognition use different resources than response (Wickens, 2008). The second dimension called codes of processing outlines that different activities along the dimension use different resources, for instance, linguistic activity will draw on a different set of resources as compared to spatial activity. The third dimension is modalities. It is a part of perception and outlines that different perceptual modalities along the dimension use separate resources, for instance, the sense of touch uses different resources than our vision. The fourth dimensions is a part of our visual resources and comprises of focal and ambient vision, that are tasked with parsing different parts of the visual scene, for instance, focal vision is more foveal and ambient vision is more relevant to the whole visual filed (Wickens, 2008).

It is important to understand that simply being on a separate level along any of these dimensions does not guarantee that there will be no interference between tasks. For instance, if one is trying to read a book while listening to the news on the radio, both tasks will suffer interference because they are both using the linguistic resources within the codes dimension of the 4-D model (Wickens, 2008).

Since drivers in autonomous vehicles are more likely to take up secondary tasks, it is quite possible that their audio-visual senses will be busy attending to the secondary task to varying degrees as predicted by the multiple resource theory (Wickens, 2002, 2008). Haptic cues provide information through the sense of touch, which is a sensory channel independent of our hearing and vision. Multiple resource theory predicts that because haptic cues use a separate sensory channel, they are not likely to suffer interference from audio-visual secondary tasks (Wickens, 2008). Driving itself is primarily a visual activity and drivers may pick up audio-visual secondary tasks such as playing games, watching videos, texting, replying to emails, or talking on the phone when an autonomous system is driving the car (Telpaz et al., 2017; Wan & Wu, 2018a, 2018b). Thus, any information provided to drivers engaged in secondary tasks should use a sensory channel other than hearing and vision.

Prior research has provided evidence that haptic cuing can convey information to drivers without interference from audio-visual senses, with less cognitive workload compared to visual signals, and is perceived to be less annoying. Graded alerts are alerts that provide signals proportional to the degree of threat, whereas single-stage alerts provide warnings only when the threat increases beyond a pre-defined limit (Lee et al., 2004). In a study comparing alert types (graded vs. single-stage) and alert modes (haptic vs. audio) researchers found that graded alerts were trusted more than single-stage alerts and haptic alerts provided using a vibrating seat were perceived to be less annoying than audio alerts (Lee et al., 2004). In a study comparing haptic cues provided using vibrotactors in the seat and visual navigation aids presented on a separate LCD screen next to the steering wheel to assist drivers in navigation, researchers found faster reaction times and lower workload for the haptic cues (Van Erp & Van Veen, 2001).

   Prior research has also found that drivers have lower response times for warning signals displayed using haptic cues when compared to other sensory channels. In a study comparing the effectiveness of rear-end collision warnings presented via audio, visual, and haptic modalities, researchers found that haptic warnings had the shortest mean response time and significantly shorter response times than drivers who were provided visual warning (Scott & Gray, 2008). Researchers also found that the haptic modality improved the response times of engaged and expectant drivers better than the other two modalities (Scott & Gray, 2008).

   Studies comparing haptic warnings to audio warnings for lane departure systems found haptic warnings to be more effective. In a study comparing lane departure warnings using haptic cues provided via seat vibration, an auditory rumble strip sound, and a multimodal alert combining haptics and audio, the researcher found the haptic modality had significantly faster reaction times compared to the audio only alert and to the multimodal alert (Stanley, 2006). Drivers perceived the haptic alert to be the least annoying while the multimodal alert scored highest in parameters such as trust, urgency, and appropriateness (Stanley, 2006). In another study evaluating lane departure warnings for drowsy drivers delivered using steering wheel torque, a rumble strip sound, steering wheel vibration, and a head-up display, researchers found the steering wheel vibration with steering wheel torque to be most effective lane departure alert for drowsy drivers evidenced by faster reaction times and smaller lane excursions (Kozak et al., 2006).

   Thus, we see that findings in the literature are in-line with the predictions of the multiple resource theory, and researchers have repeatedly and consistently found haptic cuing to be a good fit for displaying a wide array of informational and warning messages to the drivers using a variety of locations in the vehicle. It appears that haptic cuing deserves continued investigation as a promising HMI modality—particularly for drivers using higher levels of automation since they are even more likely to be distracted and have their audio-visual sensory channels occupied by secondary tasks (Telpaz et al., 2017; Wan & Wu, 2018a, 2018b). Continuing towards the goal of designing a solution to maintain driver situation awareness during automated driving while the driver is engaged in NDRTs, I have discussed TORs and haptic cuing as they relate to HMI, driver interaction with autonomous systems, multiple resource theory to understand how NDRTs engage drivers, and some examples of haptic cuing in the literature which provide evidence that haptic cuing has worked well for displaying various informational and warning messages to the drivers. The next part of the picture is a discussion of the haptic driver support literature to understand in further detail how haptic cuing has been implemented in driver support, the type of haptic cuing that has been implemented, the locations in the vehicle or on the driver where haptic cuing has been implemented, and the interventions investigated by the researchers.

**Haptic Driver Support**

The next piece of the puzzle that remains before discussing a solution to maintain driver situation awareness during automated driving while the driver is engaged in NDRTs is to cover in additional detail how researchers have provided haptic driver support in the past, and to use that knowledge to design an intervention, consistent with prior literature, that would be effective in maintaining driver situation awareness during automated driving while the drivers are engaged in NDRTs, without having a negative impact on their experience and without interfering with the NDRTs. Indeed, the haptic cuing used in this research was consistent with prior research in haptic driver support which has covered areas such as issuing unimodal and multimodal TORs, issuing TORs to drivers while they were engaged in secondary tasks with the autonomous system driving the car, assisting the drivers in controlling the vehicle while they were driving, displaying spatial information about the traffic close to the drivers while the autonomous system was driving the car, and displaying warning messages and informational cues to subjects engaged in visual tracking tasks as substitutes for driving.

Prior research in issuing TORs to drivers while they were engaged in NDRTs such as reading, texting, watching videos, playing games, or sleeping during autonomous driving has included a study where researchers evaluated a range of vibrotactile TOR patterns based on objective measures of takeover performance obtained from the driving simulator and subjective measures such as perceived vibration intensity, trust and acceptance, and driver workload during takeover. Researchers found that the pattern providing cues twice on the back of the seat and then twice on the seat bottom had the best response time and time to collision during takeover (Wan & Wu, 2018a). Researcher found no significant interactions between various haptic cue patterns and NDRTs. Additionally, there was no main effect of vibration pattern type on the drivers' engagement in NDRTs and no significant effects or interactions for the vibrotactile cues or the NDRTs for the subjective measures (Wan & Wu, 2018a).

In a study comparing curve negotiation performance in young drivers using a shared control setup, which allows both the driver and the support system to affect steering wheel torque, researchers found that drivers who had haptic guidance provided on the steering wheel improved the safety boundaries of their curve negotiation, with fewer and smoother steering adjustments compared to subjects who were not provided haptic guidance (Mulder et al., 2008). Following up from their study with younger drivers, researchers conducted the same experiment with elderly drivers and found haptic guidance on the steering wheel provided similar performance benefits during curve negotiation. However, researchers also identified concerns about the increase in the amount of steering forces during haptic guidance and indicated that they would further investigate and optimize the force levels presented to the drivers on the steering wheel (Mulder & Abbink., 2010). Another research study compared lane keeping behavior in drivers provided with continuous haptic feedback to drivers provided with feedback which triggered when tolerance limits were exceeded and found that continuous guidance resulted in better performance and satisfaction compared to limit-based feedback, but was also associated with aftereffects and increased variability in driver torque on the steering wheel (Sebastiaan M. Petermeijer, Abbink, & de Winter, 2015). Researchers defined aftereffects as "steering behavior that occurs after haptic feedback is suddenly disengaged" (Sebastiaan M. Petermeijer, Abbink, & de Winter, 2015).

In a study investigating driver trust by comparing the display of system situation awareness by using spatial information about the traffic close to the drivers while the autonomous system was driving the car using either directional cues or non-directional cues, researchers found that vibrotactile cues that contained directional information about traffic vehicles increased driver trust in the autonomous system more than the cues with no directional information (Sonoda & Wada, 2017). In another study that evaluated using haptic cuing from a seat to display spatial information about approaching vehicles to assist drivers in taking over control, researchers found that drivers reported high levels of satisfaction with the haptic seat and haptic cuing led to faster and more efficient responses to TORs (Telpaz et al., 2015). Researchers also reported that haptic cuing influenced drivers' visual scanning patterns such that drivers gazed more frequently towards the vehicle's mirrors immediately following a TOR (Telpaz et al., 2015).

In a literature review evaluating 70 empirical studies on haptic feedback to drivers, researchers found that warnings commonly used vibrotactile feedback whereas guidance largely took the form of continuous force feedback (Sebastiaan M. Petermeijer, Abbink, Mulder, et al., 2015). While the cues used in my research are not warnings, they were implemented as vibrotactile stimuli nonetheless to maintain consistency with how haptic feedback is usually provided in the literature. Additionally, in the same literature review, researchers found haptic feedback provided various short term benefits such as improved performance, reduced reaction times, and reduced mental workload (Sebastiaan M. Petermeijer, Abbink, Mulder, et al., 2015).

Prior to discussing the use of haptic cues to maintain driver situation awareness during automated driving while the drivers are engaged in NDRTs, it is also important to cover the locations in the vehicle and on the drivers that researchers have previously used to provide haptic cues to gain an understanding of where these cues have been provided, how successful haptic cuing was as an implementation modality for those interventions, and to illustrate that the location of haptic cuing used in this research was consistent with previous literature. In prior research, studies have provided haptic feedback for driver support in the seats (S.M. Petermeijer et al., 2017; Telpaz et al., 2017, 2015; Wan & Wu, 2018a, 2018b; Yun & Yang, 2020), beneath the seats (Cao et al., 2010), seatbelts (Adell et al., 2008; Chun et al., 2012, 2013; Scott & Gray, 2008), wrists (Politis et al., 2014; Sonoda & Wada, 2017), upper arms (Straughn et al., 2009), waists (Politis et al., 2013), steering wheels (de Nijs et al., 2014; Melman et al., 2017; Mulder et al., 2012, 2008; Mulder & Abbink., 2010; Mulder & Abbink, 2011; Sebastiaan M. Petermeijer, Abbink, & de Winter, 2015; Tsoi et al., 2010), gas pedals (Adell et al., 2008; Mulder et al., 2010, 2011), and joysticks (Flemisch et al., 2008).

In a study studies where haptic cues have been provided in seats, researchers compared TORs that comprised of static and dynamic directional vibrotactile cues for correct response rate, reaction times, and eye and head orientation. They found that static cues performed better than directional cues, with steering wheel touch and input reaction times being lower for static patterns (S.M. Petermeijer et al., 2017). The researchers also found that head and eye movement responses were quicker for statis cues compared to dynamic cues (S.M. Petermeijer et al., 2017). In another study that provided haptic cues in the seat, researchers investigated the effectiveness of cues that displayed spatial information about vehicles approaching from behind to improve driver preparedness prior to transfer of control. Researchers found that the cues improved speed and efficiency of reactions when lane changes were required immediately after transition to manual control (Telpaz et al., 2017). In the same study, the researchers also analyzed eye tracking data and found that drivers had more systematic scan patterns of the mirrors immediately following a TOR (Telpaz et al., 2017).

Among studies where haptic cues were provided on the seatbelt, in a study designed to test eight unimodal and multimodal HMI alternatives for informing and warning the drivers for safe speed, safe distance, and the current speed limit, the researchers found the haptic cues that combined force feedback on the gas pedal and seatbelt vibrations resulted in the lowest proportion of time spent in an unsafe situation, improved driver reaction time more than other HMI modalities, and received the most number of positive ratings from the drivers (Adell et al., 2008). The visual HMI alternatives also received positive ratings whereas the audio feedback was least well received (Adell et al., 2008). In a set of two studies comparing vibrotactile feedback provided on the seatbelt and the steering wheel for forward collision warning and blind spot warning, researchers found lower reaction times and higher collision prevention rates with haptic warnings for forward collision warning. They also found the highest collision prevention rates, longest minimum distance for collision avoidance, and the highest driver preference for the blind spot warnings delivered via the steering wheel (Chun et al., 2012, 2013).

In a study comparing the effect of stimulus-response compatibility for tactile, auditory, and no warning conditions, researchers provided cues on drivers' left and right biceps using a single tactor and Velcro straps (Straughn et al., 2009). They found that incompatible warnings were best for early warnings and compatible warnings were best for late warnings. Additionally, reaction times for early warnings were significantly shorter with tactile cues (Straughn et al., 2009). The researchers note that the relationship between stimulus-response compatibility and collision warning effectiveness depends on whether the warnings occur early enough to allow drivers to evaluate the situation prior to a possible collision (Straughn et al., 2009).

In prior research investigating haptic feedback on steering wheels, researchers compared manual control with haptic shared control (Mulder et al., 2012). They found that haptic shared control improved driver performance compared to manual control by reducing the amount of steering wheel control activity needed to improve safety performance. The researchers also found that full automation improved safety further without the need for human control, but changed the task from active involvement into supervision (Mulder et al., 2012). They note that haptic shared control not only improved driver performance but also kept drivers in the loop by keeping them actively involved in the task, thereby addressing some of the challenges associated with full automation (Mulder et al., 2012). In another study where researchers investigated properly functioning and malfunctioning haptic shared control with feedback provided on the steering wheel, researchers found that for a 1.4s time-to-contact, haptic shared control reduced the hit rate to 15.2% compared to 21.2% during manual control (Mulder & Abbink, 2011). Additionally, when the shared control malfunctioned, the hit rate for 1.4s time-to-contact was increased to 64.7%, which was much higher than the hit rate during manual control, but substantially lower than the 100% hit rate with full automation (Mulder & Abbink, 2011). In a study which evaluated haptic shared control for lane keeping during lane changes with haptic feedback provided on the steering wheel, researchers found that though haptic shared control increased the initial steering wheel torque at the beginning of the lane change, the lane keeping performance was improved compared to manual control and drivers felt comfortable and in control during lane changes with haptic shared control engaged (Tsoi et al., 2010).

Among studies where haptic feedback was provided on gas pedals, researchers designed a haptic gas pedal to assist drivers during car following. The researchers addressed three design issues for the haptic gas pedal which included quantifying the separation between the vehicles, the type of haptic feedback to be provided, and the relation between vehicle separation and the haptic feedback (Mulder et al., 2011). They found that the haptic gas pedal improved drivers' vigilance car following and did not increase their workload (Mulder et al., 2011).

In prior research covering haptic feedback on joysticks, researchers investigated varying levels of haptic guidance, from manual to full automation, on a joystick during a driving task that involved driving on a two-way curvy road. They included a secondary visual search task on a separate monitor in the vehicle and a condition with automation failure (Flemisch et al., 2008). Based on preliminary data collected from 10 subjects, researchers found the haptic device was rated as being useful, easily understandable, and safe compared to driving manually with the joystick (Flemisch et al., 2008). Additionally, the researchers found that drivers completed a higher number of tasks as the level of automation increased indicating that the drivers had lower mental workloads when higher levels of automation were available (Flemisch et al., 2008). The researchers also found the driving performance to be fairly similar in all conditions except the fully automated condition, since the automation had been tuned for driving conservatively (Flemisch et al., 2008). Drivers were able to recover from automation failure in the full automation condition and were able to keep the vehicle on the road (Flemisch et al., 2008).

Thus far I have discussed previous literature covering TORs and haptic cuing as they relate to HMI, driver interaction with autonomous systems, multiple resource theory, and haptic driver support. Each of these has provided a piece of the picture which led to my decision to implement unimodal vibrotactile cues as a solution to maintain driver situation awareness during automated driving while the driver is engaged in an NDRT and without disengaging the driver from the NDRT. This research adds to the literature by using haptic cues that provided information about upcoming road and traffic conditions to update the situation awareness of the driver and without having the driver disengage from the secondary task. This approach would be less disruptive to the driver because haptic cues use a separate sensory channel independent of the audio-visual senses. Haptic cuing is likely to not suffer interference from the secondary task even if the audio-visual senses are busy attending to the secondary task.

In this research, the drivers would be engaged in a secondary task with a Level 4 autonomous system doing the driving. The haptic cues to be used corresponded to the following road and traffic scenarios:

1. Single-vehicle or multi-vehicle accident ahead
2. Traffic jam ahead
3. Road construction ahead
4. Animal on road
5. Road closed

A literature search focused on road and traffic scenarios as they relate to driving yielded no results. Therefore, the scenarios identified above were chosen because they represent a range of complexity in the driving environment and any of the five scenarios could occur during day-to-day driving. Haptic cuing was used instead of audio-visual cues because drivers in autonomous cars are likely to be engaged in secondary tasks. These tasks would occupy their audio-visual sensory channels and the drivers may not notice an audio-visual cue due to interference from the secondary task.

Also, it is critical to make the distinction that these cues are neither pre-alerts nor TORs and may or may not be followed by a TOR depending on the capabilities of the autonomous system. Pre-alerts are alerts that are issued in addition to a TOR and are followed by TORs (van der Heiden et al., 2017). In this research, the cues were not followed by a TOR. The distinction that these cues are neither pre-alerts nor TORs is important because pre-alerts and TORs, by definition, require the drivers to disengage from NDRTs to take over control from the autonomous system whereas the cues in this research are intended to provide information to the drivers without disengaging them from NDRTs. Additionally, because of this distinction, unlike pre-alerts and TORs that are intended to be disruptive, the cues used in this research are intended to have low perceived disruptiveness and were designed with that goal in mind. The goal of low perceived disruptiveness is also the reason why these cues are unimodal and limited to using the sense of touch only as opposed to being a part of a multimodal stimulus since multimodal stimuli are perceived to be more annoying (Politis et al., 2013).

## Chapter 2: Phase 1 – Cue and Tactor Selection
### Method

This research comprised of two phases. The first phase (Phase 1) was intended to create a list of haptic cues that would be used to inform the subjects about upcoming road scenarios and identify a device that could be used to display the haptic cues. The second phase (Phase 2) used the haptic cues and the device selected based on the results of Phase 1 to display haptic cues to maintain the situation awareness of drivers during automated driving while the drivers were engaged in a secondary task and without disengaging them from the secondary task. Subjects played Fruit Ninja on a smartphone as their secondary task during both phases. A game was chosen as the secondary task because studies have shown that subjects have a higher task engagement when playing a game on a smartphone versus reading, typing, and napping (Wan & Wu, 2018b). In Phase 1, C2 and C3 tactors (see Figure 1) were used to compare seven haptic cues to determine whether C2 or C3 tactors would have a higher cue recognition accuracy along with higher satisfaction ratings, and to identify which five cues out of seven had the highest cue recognition accuracy.

### Subjects

There were 14 subjects recruited for Phase 1. Subjects were recruited from the Rice University undergraduate subject pool and were compensated for their participation with credit towards a course requirement. There were 12 female and 2 male subjects with their age ranging from 18 to 20 years ($M = 18.93$ years).

### Design

To help determine whether C2 or C3 tactors would have a higher cue recognition accuracy along with higher satisfaction ratings and to identify which five cues out of seven had the highest cue recognition accuracy, a within-subjects design with random assignment was used in Phase 1 with all subjects performing the cue recognition test on both C2 and C3 tactors in a randomized order. Subjects were trained on the seven haptic cues prior to the recognition tests. Fruit Ninja was used as the secondary task during cue recognition. Cues were presented to subjects in Latin square to counterbalance for any order effects. Each cue was given a name to help subjects with the recognition task. Subjects were informed that the cue names were provided to help them in the recognition task and may not be representative of any characteristic of the cue they are associated with.

**Figure 1**
*Engineering Acoustics (EA) C3 Tactor (Left) and C2 Tactor (Right). C2 Tactors are 1.2 inches in diameter, 0.31 inches in height, and weigh 17 grams. C3 Tactors are 0.8 inches in diameter, 0.25 inches in height, and weigh 8 grams (Engineering Acoustics Inc., n.d.).*

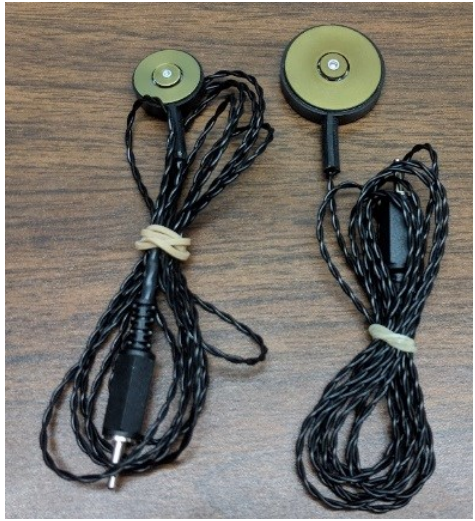

**Materials and Stimuli**

A 2016 Macbook Pro, a haptic device made of two C2 tactors and a Gymboss exercise armband, a second haptic device made of two C3 tactors and a Gymboss exercise armband, and one Syntacts Amplifier board and its API (Pezent et al., 2020) were used to display cues by attaching the tactors to the subject's upper arm. Tactors used in both devices were kept 4.5 inches apart, measured from the center of one tactor to the center of the second tactor. Syntacts GUI was used to design all haptic cues (Pezent et al., 2020). Haptic cue names and design specifications are shown in Table 1, and sample Python scripts to play the seven haptic cues and their visualizations in Syntacts GUI are provided in Appendix A. A Google Nexus 6P with Fruit Ninja installed was provided to the subjects so they could play the game as their secondary task. A survey with the System Usability Scale (SUS), a question about cue disruptiveness, and an area for additional comments was created using Google Forms. To evaluate cue disruptiveness, subjects were asked to rate the statement "I felt the vibration feedback was disruptive" on a scale of 1 to 5 with 1 being strongly disagree and 5 being strongly agree. The cue recognition test was also administered using Google Forms. Additional materials included Personal Protective Equipment (PPE) and cleaning supplies to lower the risk of infections while collecting data in person during the COVID-19 pandemic.

**Table 1**
*Haptic Cue Specifications*

| Cue Name | Cue Specifications |
| --- | --- |
| Cue 1: Jackhammer | 200Hz Sine wave, duration = 1.5s, amplitude = 0.5, pulse-width modulated at 10Hz with a 0.3 duty cycle and scalar addition of 1 presented simultaneously on both tactors. |
| Cue 2: Laser | 200Hz Square wave, duration = 0.5s, amplitude = 0.4 presented once on Tactor 1 then on Tactor 2 after a 0.05s wait. |
| Cue 3: Bunny Hop | 256Hz Square wave, duration = 0.05s, amplitude = 0.4 presented 4 times on Tactor 1 then on Tactor 2 after a 0.05s wait. |
| Cue 4: Passing By | 10Hz Sine wave, pulse-width modulated at 230Hz with a 0.5 duty cycle and an ASR of .25s 1s .25s, amplitude = 0.4 presented once on Tactor 1. |
| Cue 5: Oncoming | First part as a 200Hz Square wave, duration = 1s, amplitude = 1 with a ramp modification starting at 0.01 and a rate of 1. Second part as a 300Hz Sine wave, duration = .1s, amplitude = 1. Presented as a combination with the first part displayed on Tactor 1 and the second part displayed on Tactor 2 with a 0.05s wait in between. |
| Cue 6: Heartbeat | First half-beat as a 440Hz Sine wave Attack Sustain Release (ASR) of .1s .15s and .1s, and second half-beat at 0.75 times the first half-beat ASR with a .1s gap in between the two combined to create one full heartbeat. Presented on repeat for 1.75s on each tactor with a .1s delay between the presentations. |
| Cue 7: Ping Pong | 256Hz Square wave, duration = 0.05s, amplitude = 0.4 presented 5 times alternating between Tactor 1 and Tactor 2 with a 0.05s wait in between. |

**Procedure**

Subjects were seated next to the computer station after obtaining informed consent. One of the two haptic devices was chosen at random and attached to the upper area of the arm the subject was not going to use to play Fruit Ninja. Subjects were trained for 10 minutes on the seven haptic cues. Once the training was over, subjects were given a chance to ask any questions prior to starting the recognition test.

**Figure 2**
*Fruit Ninja Classic Mode Gameplay*

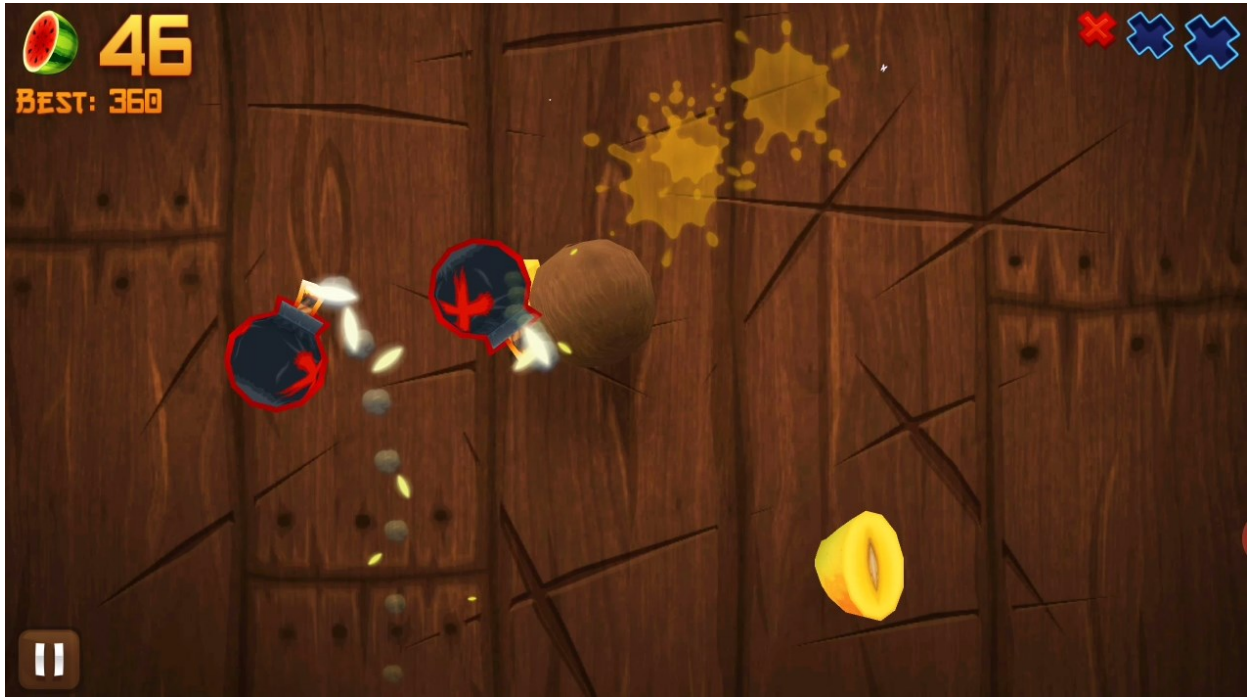

      Subjects were asked to start playing Fruit Ninja in Classic Mode (see Figure 2), and haptic cues were displayed to them using the haptic device in Latin square. Fruit Ninja is a popular smartphone game where players earn points by slicing fruits while trying to avoid slicing bombs. They slice fruits by swiping across the screen with a finger. A running tally of their game score is shown on the top left of the screen and the number of lives remaining is shown on the top right of the screen. Classic Mode is oldest and the initial set of gameplay rules that Fruit Ninja launched with. In Classic Mode players lose a life if they miss slicing a fruit and the game is over once a player runs out of lives. The game is also over if a player slices a bomb instead of a fruit. Classic Mode was chosen since it is the oldest game mode in Fruit Ninja, and it was likely that all the subjects would already be familiar with that mode. Indeed, subjects in both phases were already comfortable playing Fruit Ninja in Classic Mode and did not require additional training in how to play Classic Mode. This allowed me to ensure that using secondary task engagement measures based on performance in Fruit Ninja remained valid measures of secondary task engagement and were not affected by previous exposure to a less common game mode. Choosing a mode other than Classic Mode would have required training subjects in how to play a mode they may not have seen before thereby giving an advantage to a subset of subjects who may have already been familiar with that mode.

      Subjects were asked to tell the proctor which cue they thought was played within a few seconds of receiving the cue to help minimize any chance of forgetting which cue was displayed. The proctor recorded the subject's responses on the Google Form questionnaire. Once all seven cues had been played the subjects took the survey and the whole procedure was repeated with the second haptic device. All the equipment was cleaned and sanitized for 30 minutes before and after each subject to reduce the risk of spreading COVID-19 during in-person data collection.

## Results

Phase 1 was intended to help identify whether the device using two C2 tactors or the device using two C3 tactors would have better cue recognition accuracy and better satisfaction. It was also intended to identify five out of seven haptic cues with the highest cue recognition accuracy. A confusion matrix was created to assess the cue recognition accuracy and survey data were analyzed to compare the satisfaction between both devices.

### Confusion Matrix

Two confusion matrices were created, one for each haptic device. The confusion matrix for the first device with two C2 tactors is shown in Table 2, and the second confusion matrix with two C3 tactors is shown in Table 3. Based on the confusion matrices, Cue 1 and Cue 4 had the lowest average cue recognition accuracy for both devices combined and were therefore removed from Phase 2.

**Table 2**

*Confusion Matrix for the Haptic Device with two C2 Tactors*

| Perceived Presentation | Actual Presentation | | | | | | |
|---|---|---|---|---|---|---|---|
| | Cue 1 | Cue 2 | Cue 3 | Cue 4 | Cue 5 | Cue 6 | Cue 7 |
| Cue 1 | 50 | 0 | 0 | 21.43 | 0 | 0 | 14.30 |
| Cue 2 | 0 | 92.90 | 0 | 0 | 0 | 0 | 7.14 |
| Cue 3 | 7.14 | 0 | 78.60 | 7.14 | 0 | 0 | 14.30 |
| Cue 4 | 21.40 | 7.14 | 0 | 57.14 | 0 | 0 | 7.14 |
| Cue 5 | 0 | 0 | 14.30 | 0 | 100 | 0 | 0 |
| Cue 6 | 0 | 0 | 7.14 | 0 | 0 | 85.70 | 0 |
| Cue 7 | 21.40 | 0 | 0 | 14.29 | 0 | 14.30 | 57.10 |

*Note.* Values shown are percentages and add up to 100% for each column.

The cue recognition accuracy for all seven cues combined was $M = 74.49\%$, $Md = 78.57\%$ for the device with C2 tactors and $M = 79.59\%$, $Md = 85.71\%$ for the device with C3 tactors. Cue 1 had an average recognition accuracy of 57.14% and Cue 4 had an average recognition accuracy of 60.71%. After removing Cue 1 and Cue 4 from the cue list, there was no significant difference in cue recognition accuracy for device with C2 tactors ($M = 82.86\%$, $Md = 85.71\%$) for the remaining five cues compared to the device with C3 tactors ($M = 85.70\%$, $Md = 92.90\%$) $t(8) = 0.31$, $p = .76$, and Cohen's $d = 0.20$. Among the remaining five cues, subjects confused Cue 3 and Cue 7 with each other about 7.14% of the time.

**Table 3**

*Confusion Matrix for the Haptic Device with two C3 Tactors*

| Perceived Presentation | Actual Presentation | | | | | | |
|---|---|---|---|---|---|---|---|
| | Cue 1 | Cue 2 | Cue 3 | Cue 4 | Cue 5 | Cue 6 | Cue 7 |
| Cue 1 | 64.30 | 0 | 7.14 | 28.57 | 0 | 0 | 0 |
| Cue 2 | 0 | 92.90 | 0 | 7.14 | 7.14 | 0 | 0 |
| Cue 3 | 7.14 | 7.14 | 64.30 | 0 | 0 | 0 | 0 |
| Cue 4 | 7.14 | 0 | 21.4 | 64.29 | 0 | 0 | 0 |
| Cue 5 | 0 | 0 | 0 | 0 | 85.70 | 0 | 0 |
| Cue 6 | 0 | 0 | 0 | 0 | 0 | 92.90 | 0 |
| Cue 7 | 21.40 | 0 | 7.14 | 0 | 0 | 7.14 | 92.90 |

*Note.* Values shown are percentages and add up to 100% for each column.

**Satisfaction**

The device using C2 tactors had an average SUS score of 73.39 and a median SUS score of 75. The device using C3 tactors had an average SUS score of 71.78 and a median SUS score of 73.75 (see Figure 3). In their response to the statement that they found the cues to be disruptive rated on a scale of 1 to 5, with 1 being strongly disagree and 5 being strongly agree, subjects rated the device using C2 tactors at 2.86 and the device using C3 tactors at 2.93 (see Figure 4).

**Figure 3**

*SUS Scores for Haptic Devices*

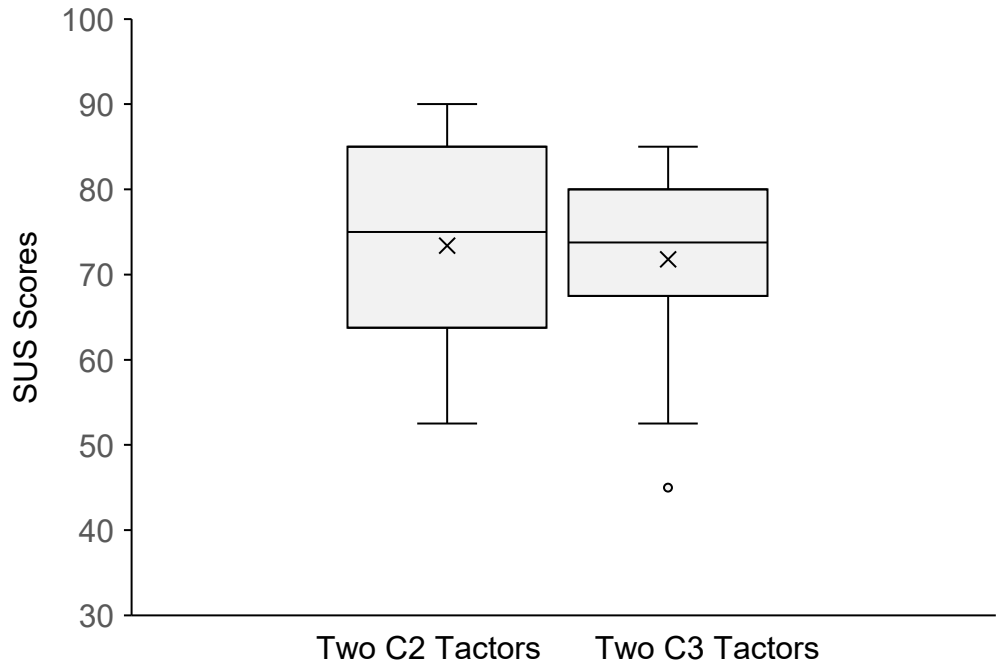

Based on their SUS scores, both devices were at an adjective rating of "good" (Bangor et al., 2009). The device using C3 tactors was selected for Phase 2 instead of the device using C2 tactors because the C3 tactors are smaller in size and lower in weight compared to the C2 tactors. Both devices had similar cue recognition accuracies and satisfaction ratings. Both devices also had similar ratings for the perceived disruptiveness of the haptic cues.

**Figure 4**
Perceived Disruptiveness of Haptic Cues (lower is better). Subjects were asked to rate the statement "I felt the vibration feedback was disruptive" on a scale of 1 to 5 with 1 being strongly disagree and 5 being strongly agree.

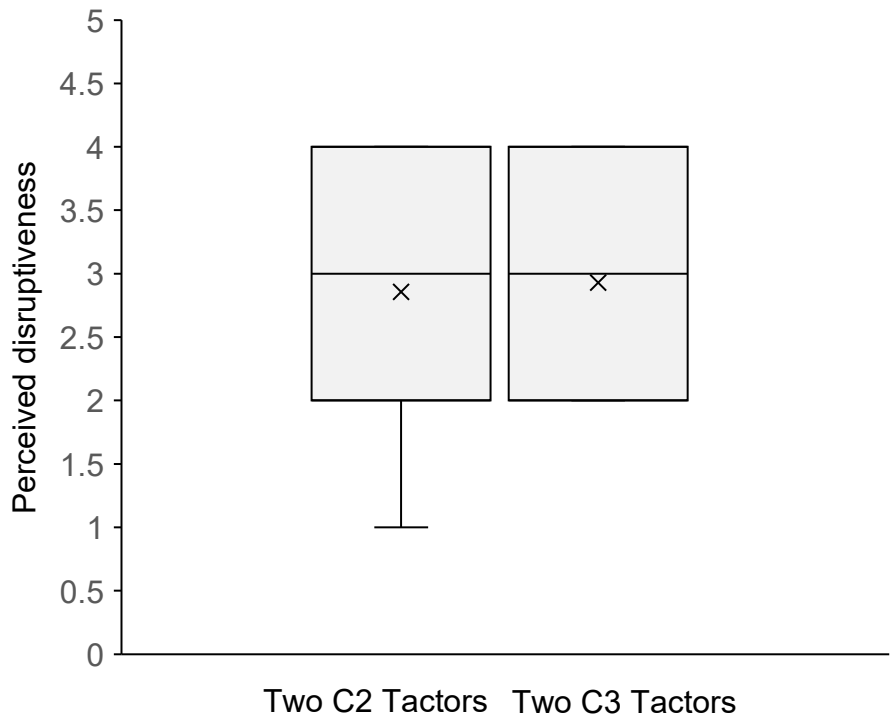

## Discussion

The device using C3 tactors was chosen since C3 tactors are smaller and lighter than C2 tactors. C2 Tactors are 1.2 inches in diameter, 0.31 inches in height, and weigh 17 grams. C3 Tactors are 0.8 inches in diameter, 0.25 inches in height, and weigh 8 grams (Engineering Acoustics Inc., n.d.). There was no evidence to suggest any difference in cue recognition accuracies, satisfaction ratings, and cue disruptiveness ratings between the two devices. Since among the remaining five cues subjects confused Cue 3 and Cue 7 with each other about 7.14% of the time it would have been ideal to redesign Cue 3 and Cue 7 to address this issue, but a decision was made to continue to Phase 2 of the data collection to complete in-person data collection for Phase 2 before the start of the flu season in Texas. It was anticipated that the impact of COVID-19 would worsen during the flu season, and there was a risk that in-person data collection would be put on hold indefinitely a second time. The decision to continue was made with the understanding that subjects may confuse Cue 3 with Cue 7 or the other way round thereby leading to less than ideal cue recognition accuracy in Phase 2.

      The reason Cue 3 and Cue 7 were confused with each other about 7.14% of the time is because they both had the same base signal, the same duration, and the same amplitude but the difference was in presentation where Cue 3 was presented four times on one tactor then four times on the second tactor whereas Cue 7 was presented 5 times alternating between the two tactors. The cues were perceptually very different to me and to most of my pilot subjects who had substantial research experience with haptics. It would have been prudent to recruit some subjects without prior haptics experience in the cue design process. This approach would have helped surface the cue design issues with Cue 3 and Cue 7 in the design phase itself.  The remaining five cues with the highest cue recognition accuracies were associated with the five road and traffic scenarios in Phase 2, one cues for each scenario.

## Chapter 3
## Phase 2 Situation Awareness - Method

In Phase 2, drivers played Fruit Ninja while seated in a driving simulator with a Level 4 autonomous system driving. A mixed design was used for the experiment in this phase with the presence of haptic cues and the SAGAT presentation time as the between-subjects conditions. Five road and traffic scenarios comprised the within-subjects part of the design. Based on the findings in Phase 1, five haptic cues with the highest cue recognition accuracies were selected for Phase 2. Additionally, the haptic device using two C3 tactors was chosen instead of the device using C2 tactors based on Phase 1 results.

**Subjects**

A total of 136 subjects were recruited for Phase 2 based on an *a-priori* power analysis in G*Power (Faul et al., 2007) for an ANOVA with a significance level of .05 comparing groups with the treatment being the presence of haptic cues and comparing them for any possible interactions dependent on when the SAGAT questions were presented. An equal number of subjects were assigned to each condition at random. The power analysis indicated that 128 subjects would be enough to detect a medium effect with 80% power (see Table 4); however, a number higher than the bare minimum required were recruited in case data from one or more subjects had to be discarded during data analysis for some reason. Subjects were recruited from the Rice University undergraduate subject pool and received credit toward a course requirement for their participation. Subjects were also required to possess a valid United States driver's license to participate. There were 75 female and 61 male subjects with their age ranging from 18 to 24 years ($M = 18.99$ years).

**Table 4**
*Power Analysis*

| Cohen's $f$ | Power | | | |
|---|---|---|---|---|
| | .8 | .85 | .9 | .95 |
| .10 | 788 | 900 | 1054 | 1302 |
| .25 | 128 | 146 | 172 | 210 |
| .40 | 52 | 60 | 68 | 84 |

**Design**

A mixed design was used for Phase 2 of the experiment with the presence of haptic cues and SAGAT presentation during vs. after as the between-subjects conditions. Subjects were randomly assigned to one of four conditions as described in Table 5:

**Table 5**
*Between-Subjects Conditions in Phase 2*

| Condition Name | Haptic Cues | Situation Awareness Questions |
|---|---|---|
| HD | Yes | During the drive |
| HA | Yes | After the drive |
| ND | No | During the drive |
| NA | No | After the drive |

Situation awareness questions about following five road and traffic scenarios comprised the within-subjects part of the design:

1. Single-vehicle or multi-vehicle accident ahead
2. Traffic jam ahead
3. Road construction ahead
4. Animal on road
5. Road closed

Subjects in cue-provided conditions were also trained on the five haptic cues and were tested for their cue recognition accuracy after the situation awareness training. Subjects in no-cue conditions were given an opportunity to memorize a list of the five scenarios in a random order that did not occur during the task. Five road and traffic scenarios were presented to subjects a in Latin square. Additionally, individual situation awareness queries associated with the scenarios were also chosen at random for each subject (Endsley, 1995b).

A variable measuring the number of scenarios identified correctly was created based on a scenario identification question at the end each time situation awareness questions were asked. Subjects in all conditions were instructed that some questions may appear to be similar, and they were to select the same response for similar questions. This was done so subjects would select the same response for some of the Level 1 situation awareness questions that were based on identifying the road conditions and the scenario identification questions itself. However, some subjects selected different responses for both questions likely because they forgot instructions provided earlier. For purposes of the variable measuring the number of scenarios identified correctly only, subjects' response was marked correct if they responded correctly to either the Level 1 situation awareness question based on identifying the road conditions or the scenario identification questions itself. Their responses to individual situation awareness and scenario identification questions were recorded without any modifications, i.e., if they responded correctly to the first question but incorrectly to the last question, then their responses to the second question was recorded as incorrect.

A variable measuring the number of correct responses per scenario was created based on the number of correct responses to the SAGAT and scenario identification questions for each scenario. A count of the number of times each subject looked up at the simulator screen during the automated drive, a tally of the game scores in Fruit Ninja, and a count of the number of games played served as objective measures of performance on the secondary task. A survey including the SUS was used to collect device satisfaction metrics in the treatment conditions. Subjects were asked to leave the smartphone on the table in front of them so that each look at the simulator screen would require a head or an eyelid movement. The proctor sat either on the left or the right side of the subject such that the proctor could clearly see every head and eyelid movement the subject made. The Look Count variable was the total number of times subjects moved either their head or their eyelids to look up at the simulator screen.

A 2x2 ANOVA was used to compare the experiment conditions using the number of times each subject looked up at the simulator screen as an inverse measure of secondary task engagement, i.e., looking up at the simulator screen fewer times indicated higher engagement in playing Fruit Ninja. The same 2x2 ANOVA was used to compare the experiment conditions using the number of games played as a measure of secondary task engagement, and to compare the experiment conditions using the mean game scores as a measure of secondary task engagement, i.e., since the skill at playing Fruit Ninja was randomly distributed in the subject pool due to random assignment, a higher mean game score would indicate higher engagement in playing Fruit Ninja.

**Materials**

Level 4 autonomous driving was simulated by using road and traffic scenario screen recordings obtained while the experimenter was driving through the scenarios in STISim Drive driving simulator. Scenario recordings were combined in Lantin square to create drive videos with autonomous driving time and time to answer situation awareness questions while the drive was paused. A computer station with two 24-inch U2417H Dell monitors running at a resolution of 1920 by 1080 pixels and a Mac Pro was used to play the drive video and display questions (see Appendix B). The Mac Pro had built-in speakers which were used to play vehicle and traffic sounds during the drives. A Logitech gaming steering wheel and pedal set was also added to the computer station to make it look like a driving simulator; however, the steering wheel and pedals were not connected for input. Auto-advancing Qualtrics questionnaires with time delays matching with drive videos were used to administer the situation awareness questions. A 2016 Macbook Pro, a haptic device made of two C3 tactors and a Gymboss exercise armband, and one Syntacts Amplifier board and its API were used to display cues by attaching the tactors to the subject's upper arm. Tactors used in the haptic device were kept 4.5 inches apart, measured from the center of one tactor to the center of the second tactor. A Google Nexus 6P with Fruit Ninja installed was provided to the subjects to play the game as their secondary task during the drive. AZ Screen Recorder app was used to record all Fruit Ninja games for each subject. A survey with the SUS, a question about cue disruptiveness, and an area for additional comments was created using Google Forms. To evaluate cue disruptiveness, subjects were asked to rate the statement "I felt the vibration feedback was disruptive" on a scale of 1 to 5 with 1 being strongly disagree and 5 being strongly agree. Additional materials included Personal Protective Equipment (PPE) and cleaning supplies to lower the risk of infections while collecting data in person during the COVID-19 pandemic.

**Procedure**

The experiment was 58 minutes in duration. Subjects were seated next to the computer station after checking their driver's license. The first 10 minutes were used for obtaining informed consent, attaching the haptic device if required to the subjects' upper area of the arm they were not going to use to play Fruit Ninja, and training subjects on how to respond to the situation awareness questions while playing Fruit Ninja in Classic Mode. A Level 4 autonomous system drove during the training and the task. Subjects played Fruit Ninja on a smartphone as their secondary task. As recommended in the literature, all subjects were trained on how to respond to the situation awareness questions prior to the task (Endsley, 1995b, 2000). Subjects were asked to not lift the phone up in the air, but rather keep the phone on the table in front of them while playing the game during the training and during the task. Subjects were told that they were free to take as many quick glances at the simulator screen as they wanted, both during the training and during the task, so long as they did not stop playing Fruit Ninja to focus entirely on the simulator screen. The situation awareness training comprised of two scenarios: school bus stopped ahead and lane closed ahead. There were four version of the situation awareness training matched with each of the four experiment conditions. Subjects in the cues provided conditions then trained on the haptic cues associated with the five road and traffic scenarios for the next 10 minutes in a random order that did not occur during the task followed by a 5-minute cue recognition test to make sure they could consistently recognize all the cues. Subjects in conditions with no cues provided were given 10 minutes to memorize a list of the five scenarios used in the task in a random order that did not occur during the task.

      Subjects spent the next 23 minutes going through a Level 4 automated drive playing Fruit Ninja in Classic Mode and answering situation awareness questions while drive was paused. The 23-minute drive comprised of 16 minutes of autonomous driving and 7 minutes to answer situation awareness questions while the drive was paused on a blank screen as recommended in the literature (Endsley, 2000). A total drive duration of 23 minutes is reasonable as prior research in driving, including automated driving, has involved drive times that can reach up to 3.5 hours and drives longer than 30 minutes in duration are quite common (Franz et al., 2015; Gugerty, 1997; Kaber et al., 2012; Rauch et al., 2008; van den Beukel & van der Voort, 2013). Subjects were given 20 seconds to respond to each query. A random delay of 1.5 minutes to 2.5 minutes was added after each traffic scenario ended and the next one began, and cues were displayed starting 3 minutes into the drive (Endsley, 1995b). Although there were no TORs in Phase 2, the subjects were told that TORs may occur at random for some subjects, and that they would need to verbalize what they would do instead of using the steering wheel to take over control from the autonomous system. They were also told to not worry if they never get a TOR during the drive.

      In the haptic cue conditions, the cues were displayed 25 seconds in advance of the road and traffic scenario to allow time for a possible TOR and the screen was blanked out 20 seconds before the road and traffic scenario. These cues are neither TORs nor pre-alerts. These cues were also not followed by a TOR. Prior research has shown that between 7-12 seconds are needed to gain situation awareness after a TOR is issued, with up to 20 seconds needed to estimate relative vehicle speeds in the surrounding traffic (Eriksson & Stanton, 2017a; Lu et al., 2017; Naujoks et al., 2018; Vogelpohl et al., 2018; Wan & Wu, 2018b). Therefore, the cues were displayed 25 seconds in advance of the road and traffic scenario. The Fruit Ninja game was also paused while the subjects responded to the situation awareness questions. In the no-cue conditions the screen was blanked out and situation awareness questions were presented one second before the vehicle took any action to address the scenario, i.e., before it started slowing down or changing lanes for instance. After the drive was over, the last 10 minutes were used to administer the survey. All the equipment was cleaned and sanitized for 30 minutes before and after each subject to reduce the risk of spreading COVID-19 during in-person data collection.

## Chapter 4
## Phase 2 Situation Awareness Results - I

In Phase 2, responses to the situation awareness questions, performance on Fruit Ninja, and survey data were analyzed to compare experiment conditions with the presence of haptic cues and SAGAT presentation time as the between-subjects conditions with the five road and traffic scenarios serving as the within-subjects part of the design. The data from two subjects was discarded because they played Fruit Ninja in Zen Mode instead of Classic Mode. Zen Mode is a different style game with different rules and scores. It is not comparable to Classic Mode. Therefore, data from the remaining 134 subjects were used to conduct the analyses.

**Number of Scenarios Identified Correctly**

A 2x2 ANOVA was used to compare the experiment conditions using the number of scenarios identified correctly by each subject for haptic cuing and for SAGAT presentation time. Boxplots comparing these conditions with number of scenarios identified correctly as the dependent variable are presented in Figure 5 and Figure 6. One subject was identified as a 3xIQR outlier; however, no outliers were removed from the data.

**Figure 5**
*Number of Scenarios Identified Correctly for the Presence of Haptic Cues. One 3xIQR outlier shown as a solid circle.*

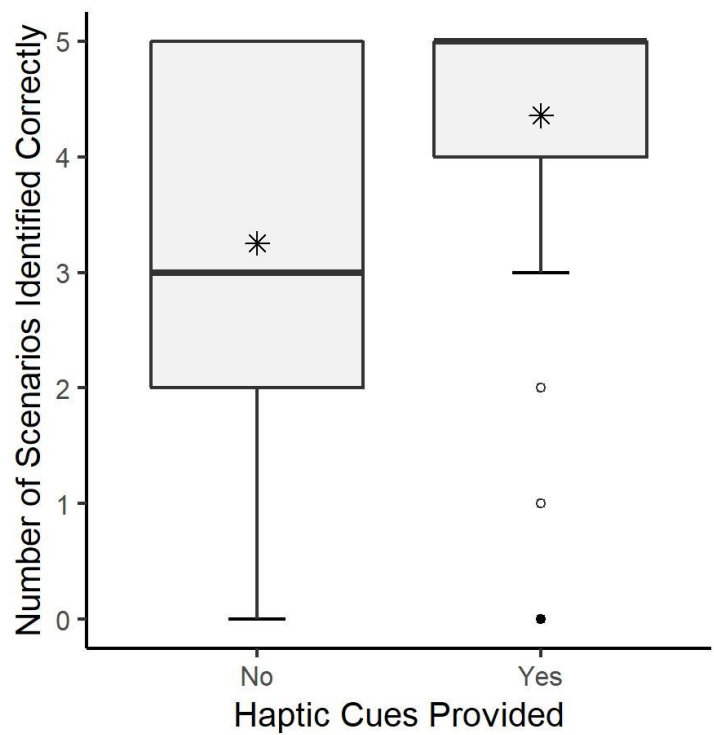

There was a significant main effect of haptic cuing $F(1, 130) = 22.73$, $MSE = 1.72$, $p < .001$, and Cohen's $f = 0.42$ with cue-proved conditions ($M = 4.33$, $Md = 5.00$) ranging from 0 to 5, and no-cue conditions ($M = 3.25$, $Md = 3.00$) ranging from 0 to 5. The main effect of SAGAT presentation time was not significant $F(1, 130) = 1.93$, $MSE = 1.72$, $p = .17$, and Cohen's $f = 0.12$ with SAGAT during the drive conditions ($M = 3.94$, $Md = 4.00$) ranging from 1 to 5, and SAGAT after the drive conditions ($M = 3.64$, $Md = 5.00$) ranging from 0 to 5. The interaction between the presence of haptic cues and SAGAT presentation time was also not significant $F(1, 130) = 0.58$, $MSE = 1.72$, $p = .45$, and Cohen's $f = 0.07$ .

**Figure 6**
*Number of Scenarios Identified Correctly for the Question Presentation Time*

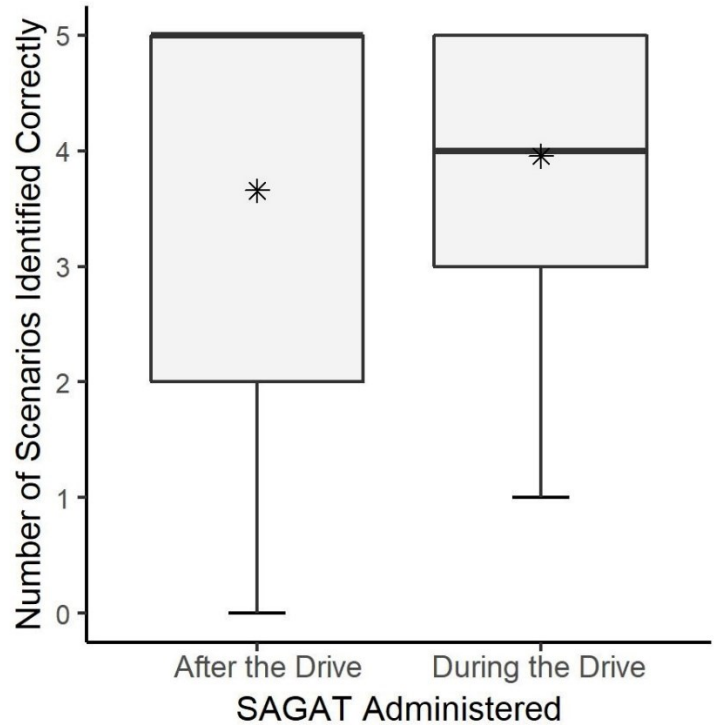

**Number of Correct Responses per Scenario**
     A 2x2x5 mixed design ANOVA was used to compare the experiment conditions using the number of correct responses per scenario for haptic cuing and for SAGAT presentation time. Bar charts comparing these conditions with number of correct responses per scenario as the dependent variable are presented in Figure 7 and Figure 8. There were 7 within-subjects outliers across all five scenarios combined and one between-subjects outlier identified using the 3xIQR criterion; however, no outliers were removed from the data. A separate analysis with the within-subjects outliers replaced and the between-subjects outlier removed is provided in Appendix C. The analysis in Appendix C has the same findings as this analysis. The number of correct responses per scenario were measured out of a maximum of 4 since there were 4 questions associated with each scenario.

**Figure 7**

*Number of Correct Responses per Scenario for the Presence of Haptic Cues. Error bars represent +/- 1 Standard Error of the Mean.*

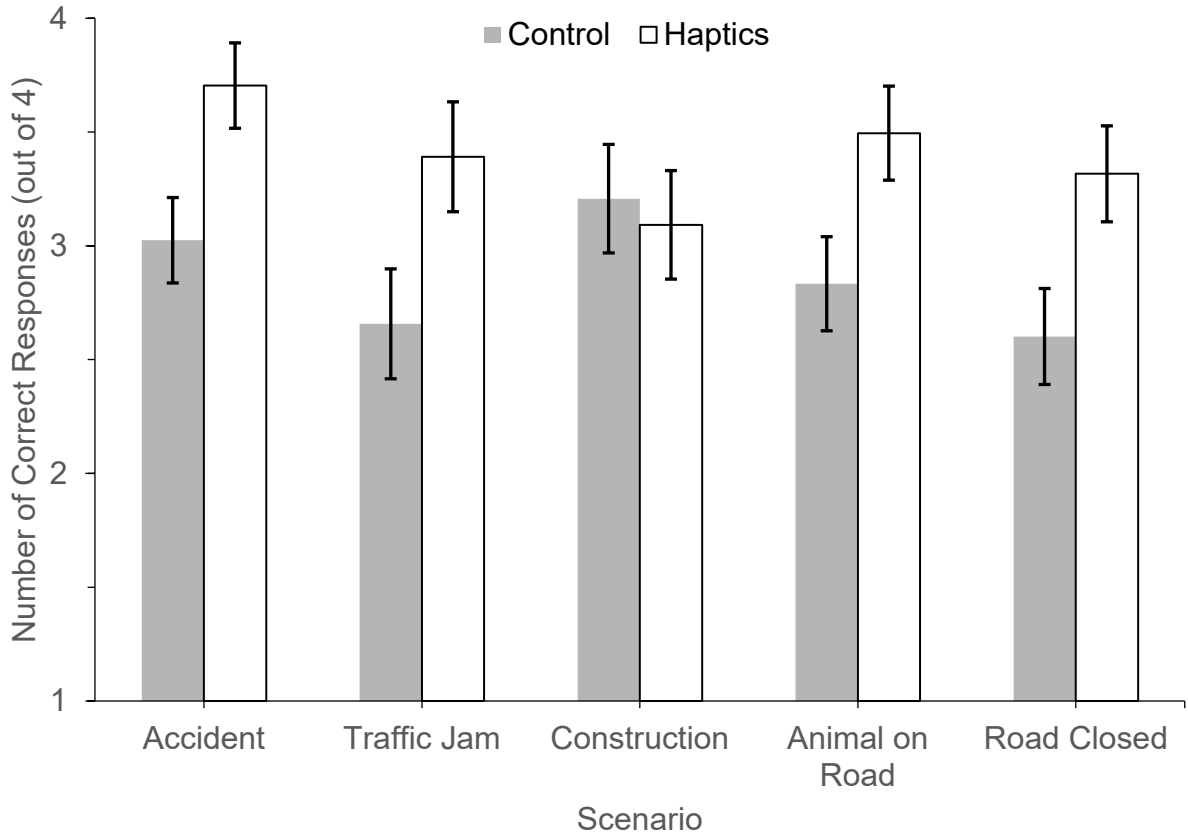

In the between-subjects part of the analysis, there was a significant main effect of haptic cuing $F(1, 130) = 15.00$, $MSE = 3.20$, $p < .001$, and Cohen's $f = 0.34$ comparing the cue-provided and the no-cue conditions. The main effect of SAGAT presentation time was not significant $F(1, 130) = 3.19$, $MSE = 3.20$, $p = .076$, and Cohen's $f = 0.16$ when comparing SAGAT after the drive conditions with SAGAT during the drive conditions. The interaction between the presence of haptic cues and the SAGAT presentation time was also not significant $F(1, 130) = 0.88$, $MSE = 3.20$, $p = .35$, and Cohen's $f = 0.08$.

**Figure 8**

*Number of Correct Responses per Scenario for Question Presentation Time. Error bars represent +/- 1 Standard Error of the Mean.*

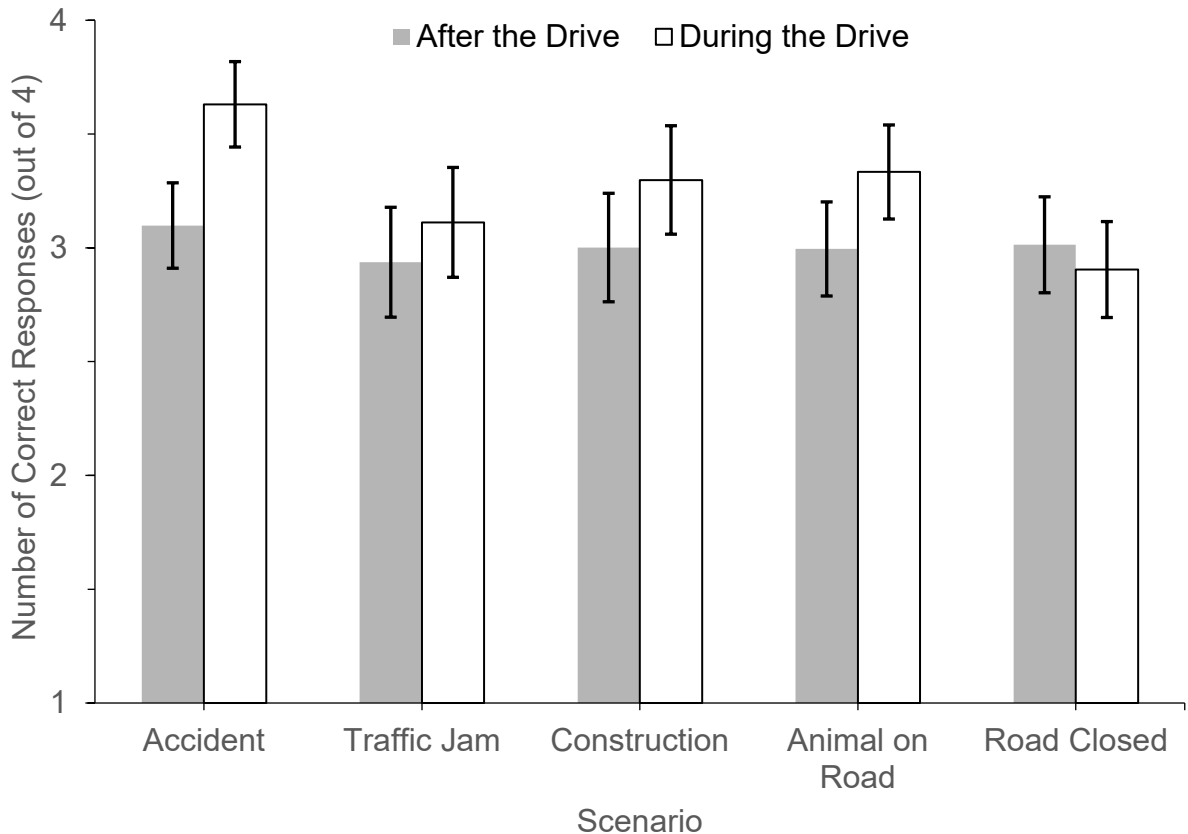

In the within-subjects part of the analysis, sphericity was assumed for the main effect of scenario type and the interactions of scenario type with the between-subjects variables as the Huynh-Feldt test reported a value greater than .95. There was a significant effect of scenario type $F(4, 520) = 2.73$, $MSE = 1.19$, $p = .029$, and Cohen's $f = 0.14$. The interaction between scenario type and haptic cuing was also significant $F(4, 520) = 3.74$, $MSE = 1.19$, $p = .005$, and Cohen's $f = 0.17$. The interaction between scenario type and SAGAT presentation time was not significant $F(4, 520) = 1.58$, $MSE = 1.19$, $p = .18$, and Cohen's $f = 0.11$. The three-way interaction between scenario type, haptic cuing, and SAGAT presentation time was also not significant $F(4, 520) = 1.94$, $MSE = 1.19$, $p = .10$, and Cohen's $f = 0.12$.

The interaction between scenario type and haptic cuing was decomposed further with an interaction contrast comparing the Road Construction scenario with the other four scenarios. This contrast was chosen because the Road Construction scenario was associated with Cue 3 and also based on Phase 1 findings which indicated that subjects confused Cue 3 and Cue 7 with each other about 7.14% of the time and based on comments from some subjects in Phase 2 that they found the cue for the Road Construction scenario and the cue for the Animal on Road scenario (associated with Cue 7) to be similar. This interaction contrast would help uncover whether the number of correct responses were indeed fewer in the Road Construction scenario as compared to the other four scenarios.

The interaction contrast comparing the number of correct responses in the Road Construction scenario with the other four scenarios was significant $F(1, 130) = 12.79$, $MSE = 27.65$, $p < .001$, and Cohen's $f = 0.31$. This tells us that there were fewer number of correct responses in the Road Construction scenario as compared to the other four scenarios. It is likely that the fewer number of correct responses in the Road Construction scenario were due to the cue design issues associated with Cue 3.

**Phase 2 Situation Awareness Results - II**
**Correct Responses per Scenario for each Level of Situation Awareness**

Fisher's Exact Test was used to compare the experiment conditions using the proportion of correct responses by the subjects per scenario for each level of situation awareness for haptic cuing and for SAGAT presentation time. Fisher's Exact Test was chosen because it makes no assumptions of the data and is designed to test differences in proportions between two variables, for instance, comparing the proportion of correct responses to situation awareness Level 1 queries between the two cue-provided conditions. Logistic regression was not appropriate because the situation awareness query data did not conform to the required binomial assumption. Even though queries were presented 30 to 80 times over the entire experiment in line with Endsley's recommendations of presenting each query at least 30 times over the entire experiment, some queries that were presented close to 30 times ended up with fewer than 10 data points per experiment condition (Endsley, 2000). Additionally, the denominators for individual SAGAT query proportions were different than the maximum value for the number of correct responses per scenario variable since each SAGAT query was selected randomly and thus had a different number of presentations across experiment conditions.

False Discovery Rate (FDR) was used to adjust the significance level for multiple comparisons (Benjamini & Hochberg, 1995). FDR-adjusted significance tests for the proportion of correct responses per scenario for each level of situation awareness are presented in this section, and the FDR-adjusted significance tests for the proportion of correct responses for each situation awareness question for each level of situation awareness are provided in Appendix C.

*Responses to Level 1 SAGAT Questions*

Bar charts comparing haptic cuing and the SAGAT presentation time for the proportion of correct responses per scenario for situation awareness Level 1 are presented in Figure 9 and Figure 10.

**Figure 9**
*Proportion of Correct Responses per Scenario for Situation Awareness Level 1 for the Presence of Haptic Cues. Error bars represent a 95% Confidence Interval of the Mean. Significant Differences marked with an \*. Chance performance shown as a dashed line.*

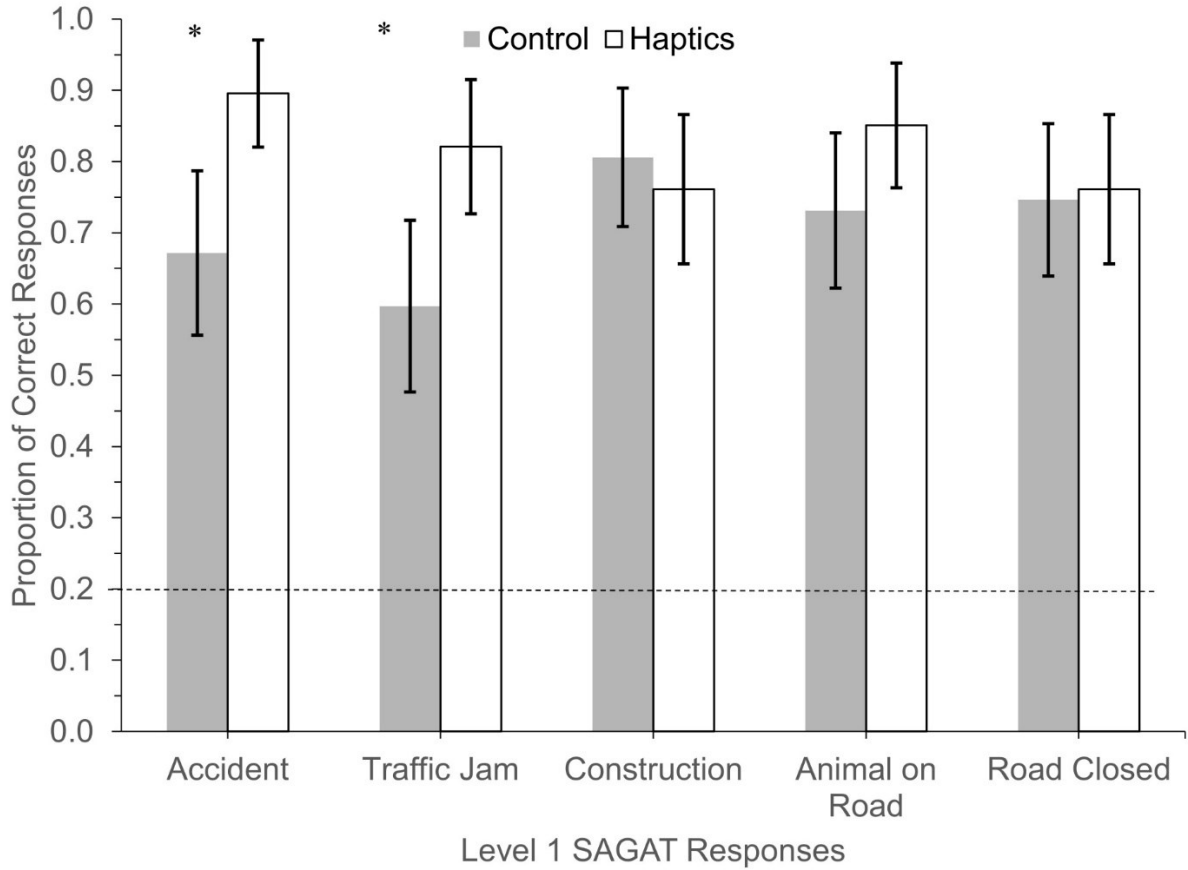

**Figure 10**

*Proportion of Correct Responses per Scenario for Situation Awareness Level 1 for the Question Presentation Time. Error bars represent a 95% Confidence Interval of the Mean. Chance performance shown as a dashed line.*

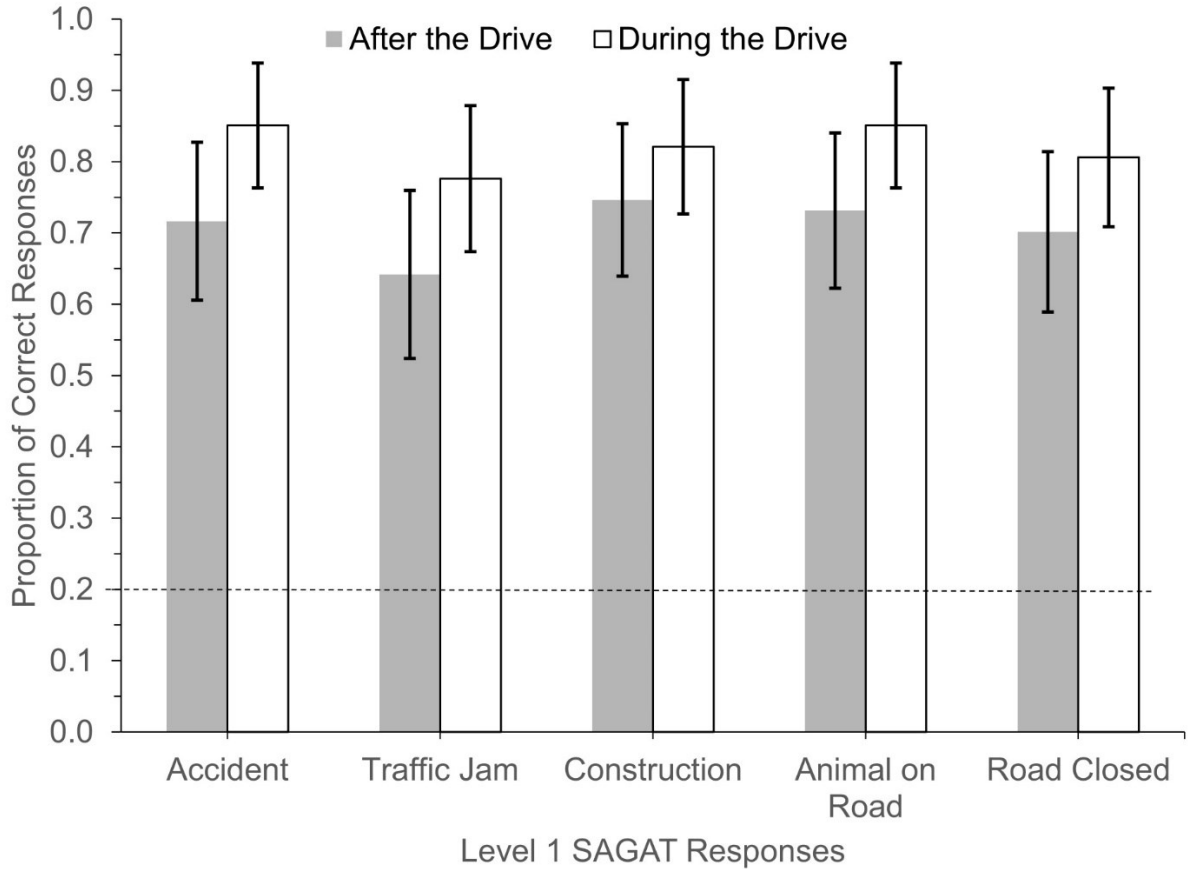

FDR-adjusted significance tests for situation awareness Level 1 for all five scenarios comparing them for haptic cuing and for the SAGAT presentation time are shown in Table 6.

**Table 6**

*Significance Tests for the Proportion of Correct Responses for Situation Awareness Level 1*

| Scenario Name | Haptics ($p$-values) | Question Presentation Time ($p$-values) |
|---|---|---|
| Accident | **.003** | .092 |
| Traffic Jam | **.007** | .13 |
| Road Construction | .68 | .40 |
| Animal on Road | .14 | .14 |
| Road Closed | > .99 | .23 |

*Note*. $p$-values significant after FDR adjustment are presented in bold.

*Responses to Level 2 SAGAT Questions*

Bar charts comparing haptic cuing and the SAGAT presentation time for the proportion of correct responses per scenario for situation awareness Level 2 are presented in Figure 11 and Figure 12.

**Figure 11**

*Proportion of Correct Responses per Scenario for Situation Awareness Level 2 for the Presence of Haptic Cues. Error bars represent a 95% Confidence Interval of the Mean. Significant Differences marked with an \*. Chance performance shown as a dashed line.*

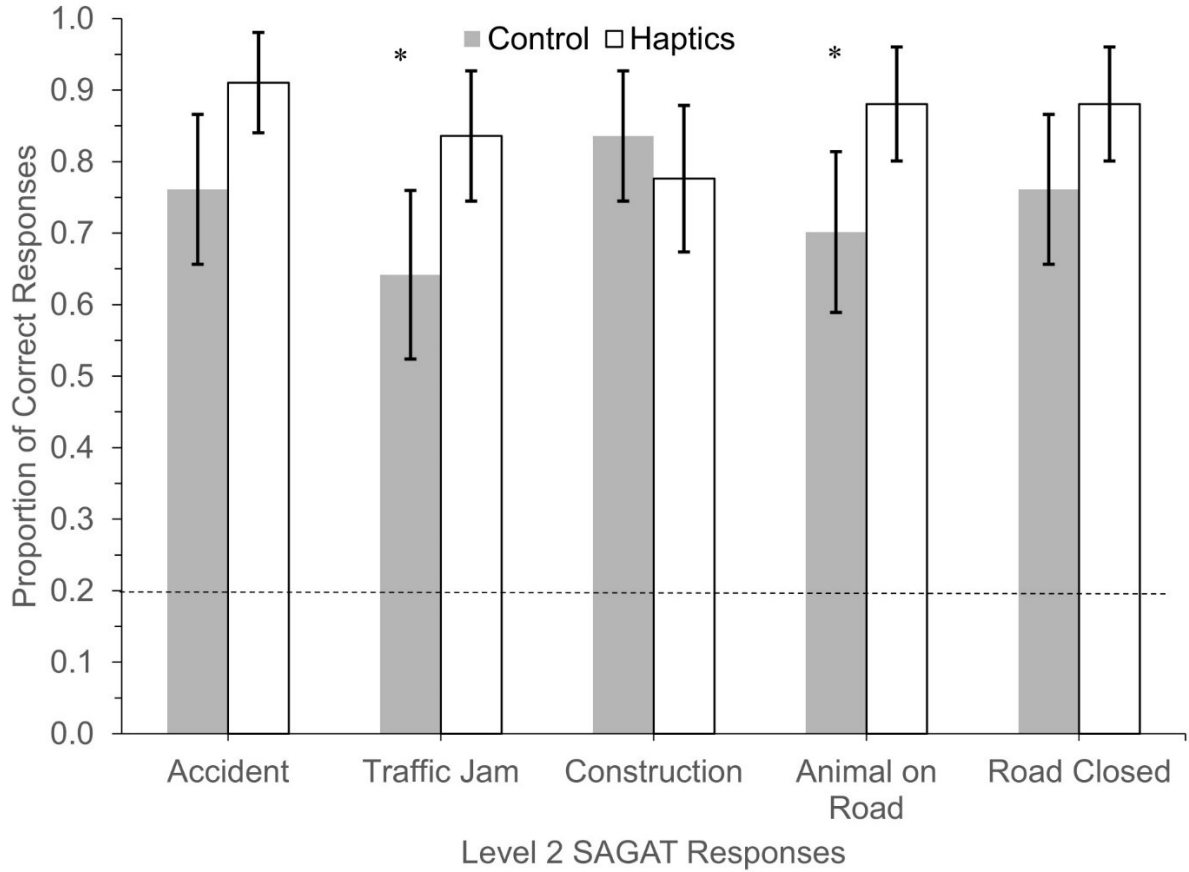

**Figure 12**
*Proportion of Correct Responses per Scenario for Situation Awareness Level 2 for Question Presentation Time. Error bars represent a 95% Confidence Interval of the Mean. Chance performance shown as a dashed line.*

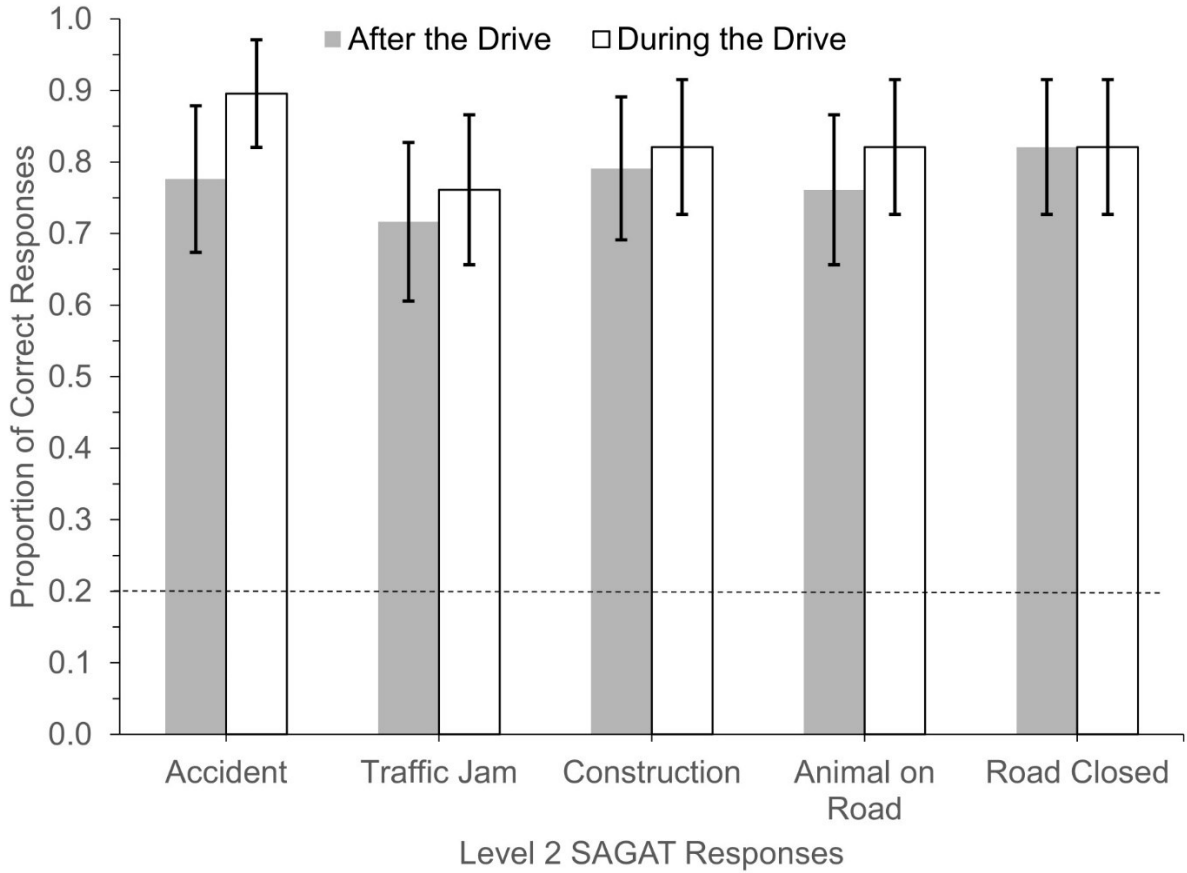

FDR-adjusted significance tests for situation awareness Level 2 for all five scenarios comparing them for haptic cuing and the SAGAT presentation time are shown in Table 7.

**Table 7**
*Significance Tests for the Proportion of Correct Responses for Situation Awareness Level 2*

| Scenario Name | Haptics (*p*-values) | Question Presentation Time (*p*-values) |
| --- | --- | --- |
| Accident | .034 | .10 |
| Traffic Jam | **.018** | .69 |
| Road Construction | .51 | .83 |
| Animal on Road | **.018** | .52 |
| Road Closed | .11 | > .99 |

*Note*. *p*-values significant after FDR adjustment are presented in bold.

*Responses to Level 3 SAGAT Questions*

Bar charts comparing haptic cuing and the SAGAT presentation time for the proportion of correct responses per scenario for situation awareness Level 3 are presented in Figure 13 and Figure 14.

**Figure 13**
*Proportion of Correct Responses per Scenario for Situation Awareness Level 3 for the Presence of Haptic Cues. Error bars represent a 95% Confidence Interval of the Mean. Significant Differences marked with an *. Chance performance shown as a dashed line.*

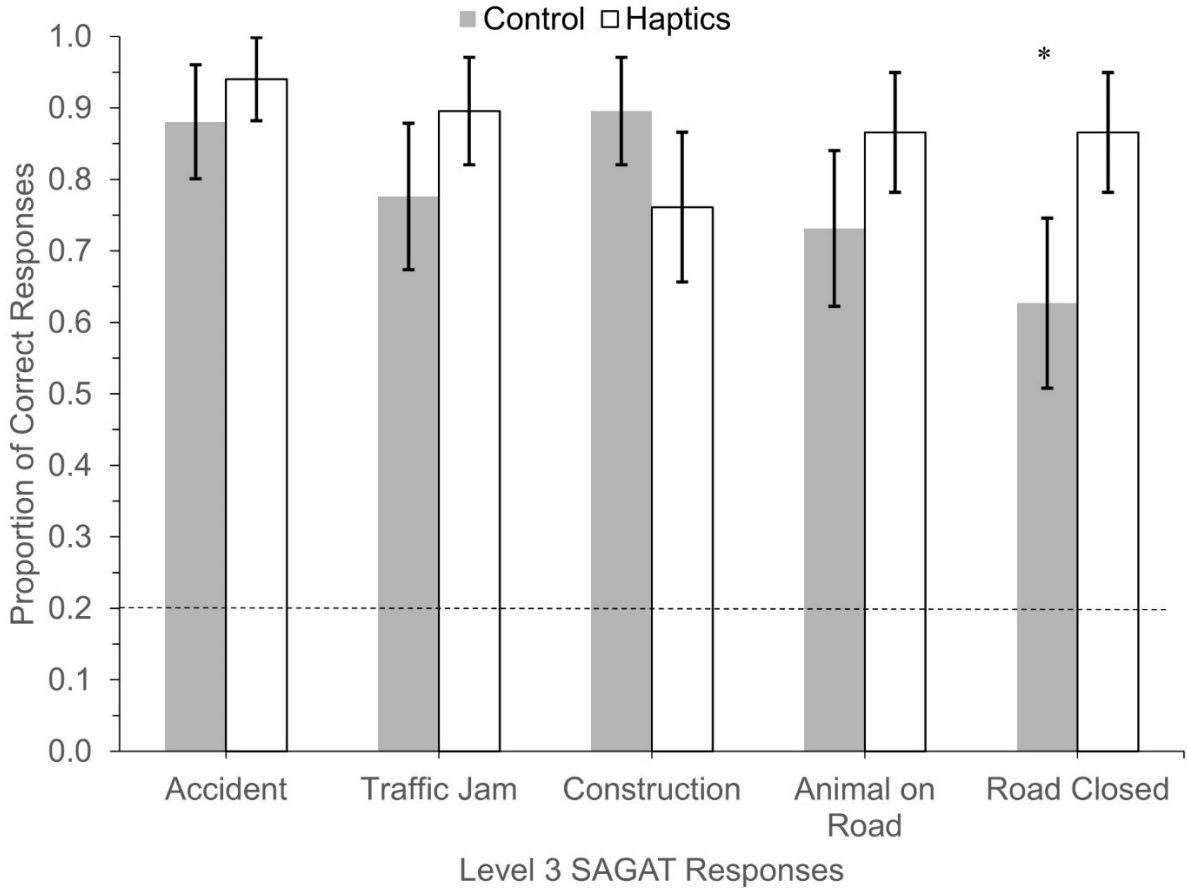

**Figure 14**

*Proportion of Correct Responses per Scenario for Situation Awareness Level 3 for the Question Presentation Time. Error bars represent a 95% Confidence Interval of the Mean. Significant Differences marked with an \*. Chance performance shown as a dashed line. The error bar for the Accident scenario and questions presented during the drive is cut off because proportion only goes up to 1.*

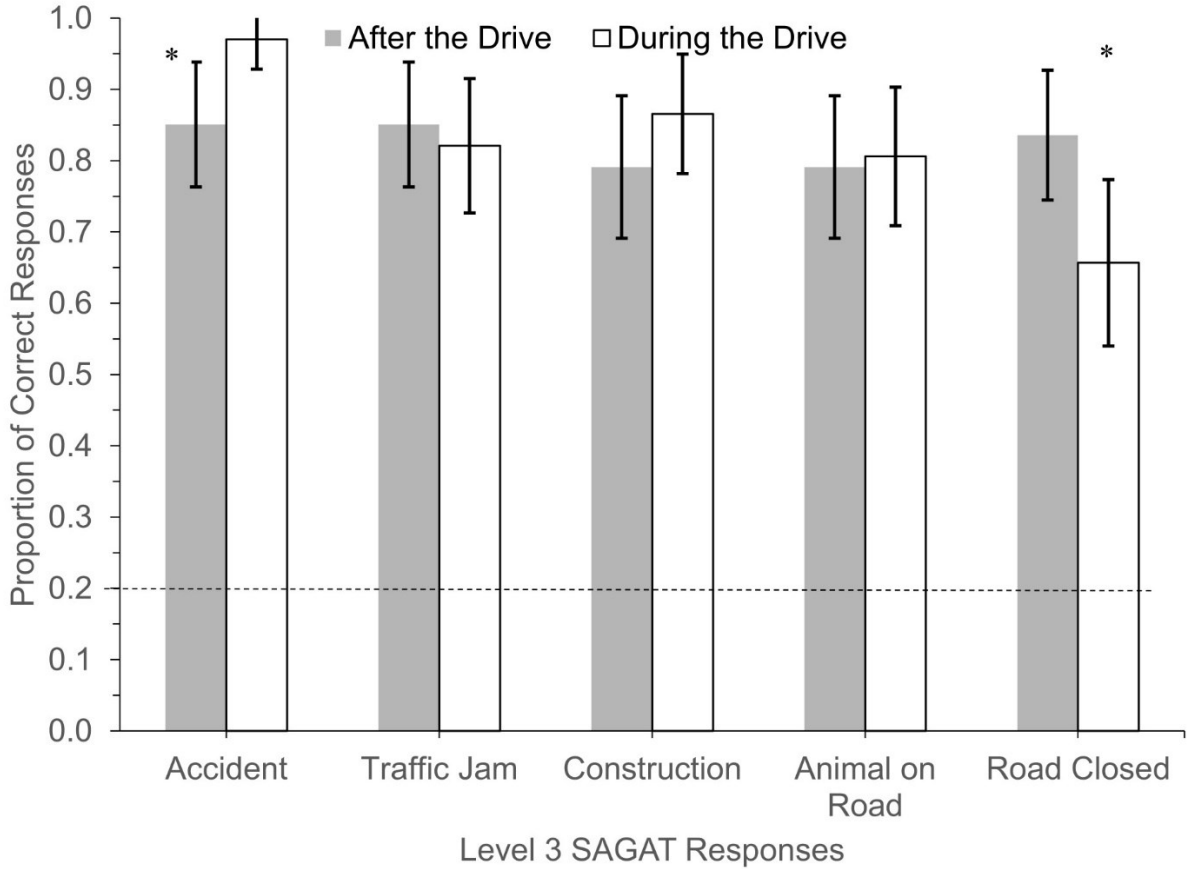

FDR-adjusted significance tests for situation awareness Level 3 for all five scenarios comparing them for haptic cuing and for the SAGAT presentation time are shown in Table 8.

**Table 8**

*Significance Tests for the Proportion of Correct Responses for Situation Awareness Level 3*

| Scenario Name | Haptics ($p$-values) | Question Presentation Time ($p$-values) |
|---|---|---|
| Accident | .36 | **.03** |
| Traffic Jam | .10 | .81 |
| Road Construction | .065 | .36 |
| Animal on Road | .084 | > .99 |
| Road Closed | **.003** | **.028** |

*Note*. $p$-values significant after FDR adjustment are presented in bold.

*Responses to Scenario Identification Questions*

    Bar charts comparing haptic cuing and the SAGAT presentation time for the proportion of correct responses per scenario for the scenario identification questions are presented in Figure 15 and Figure 16.

**Figure 15**
*Proportion of Correct Responses per Scenario for the Scenario Identification Questions for the Presence of Haptic Cues. Error bars represent a 95% Confidence Interval of the Mean. Significant Differences marked with an \*. Chance performance shown as a dashed line. The error bar for the Accident scenario and questions presented during the drive is cut off because proportion only goes up to 1.*

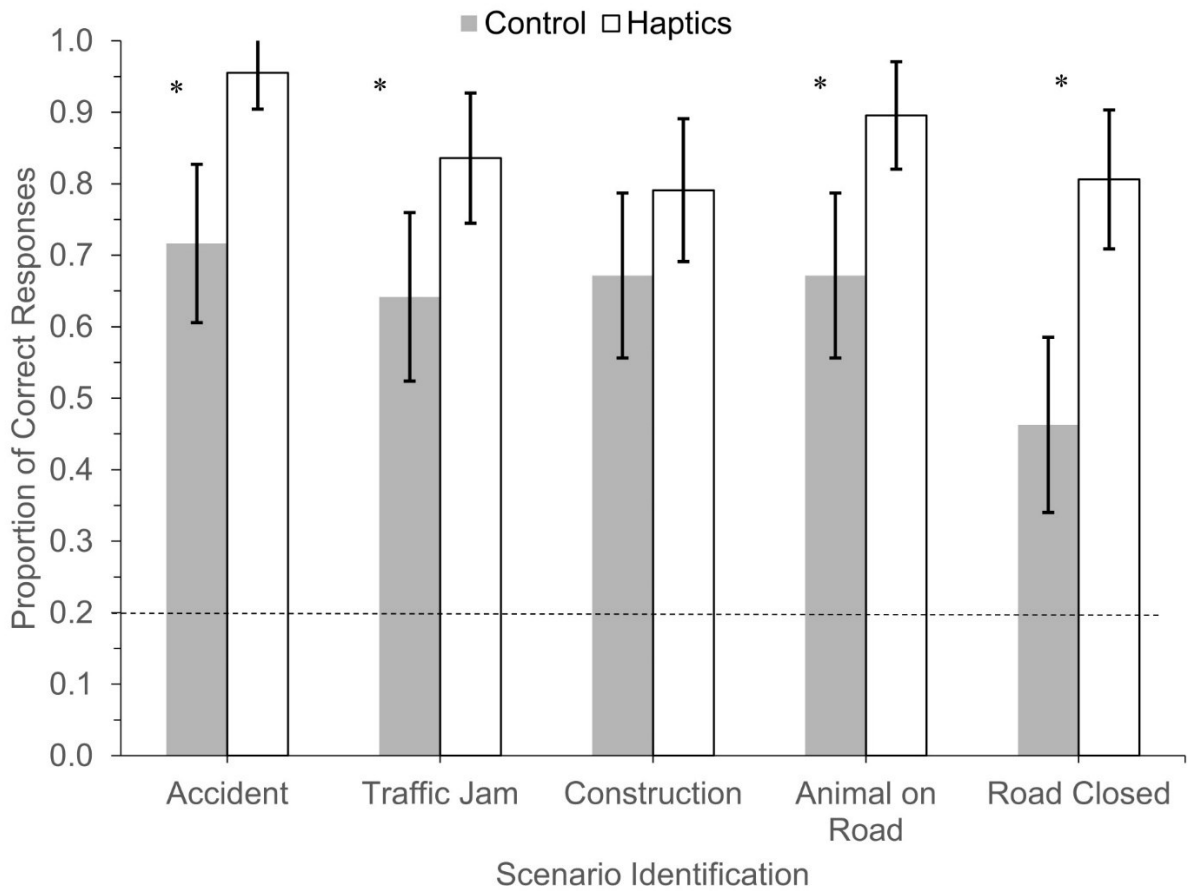

**Figure 16**

*Proportion of Correct Responses per Scenario for the Scenario Identification Questions for the Question Presentation Time. Error bars represent a 95% Confidence Interval of the Mean. Significant Differences marked with an *. Chance performance shown as a dashed line.*

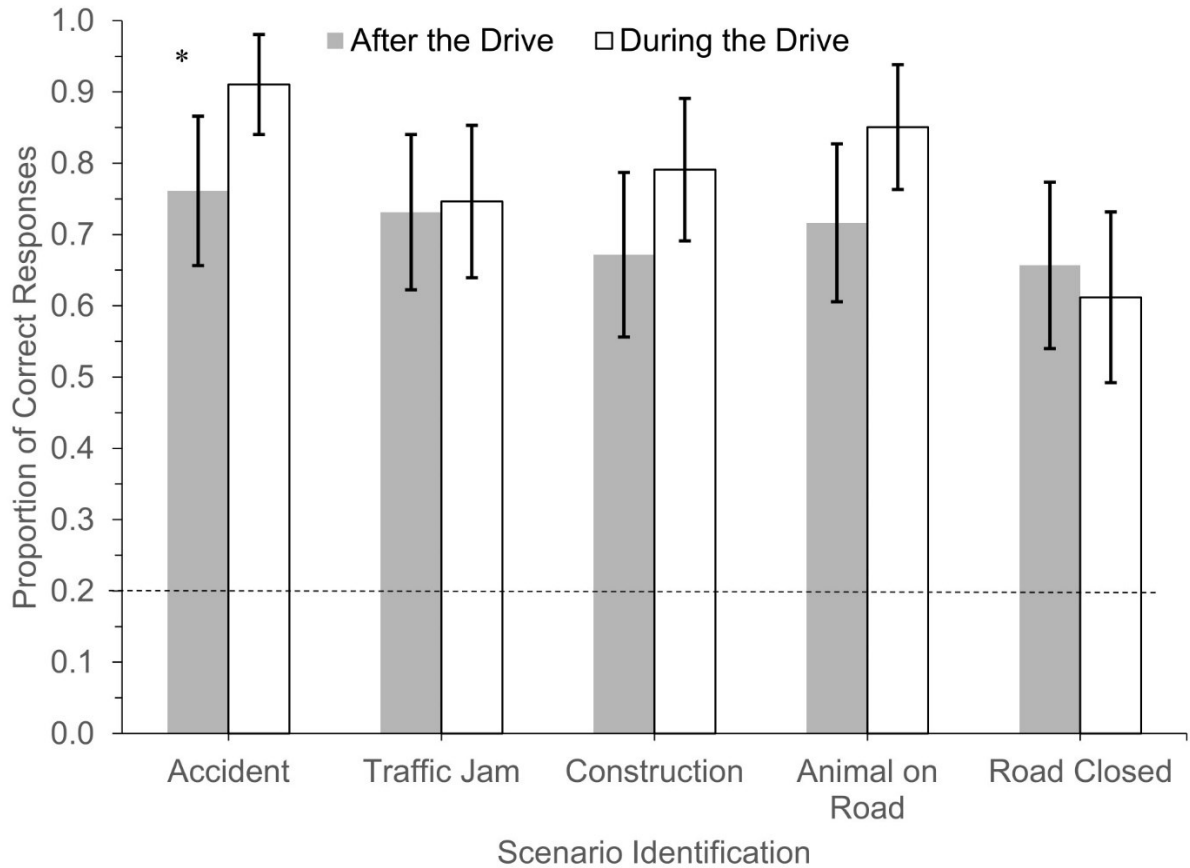

FDR-adjusted significance tests for the scenario identification question for all five scenarios comparing them for haptic cuing and for the SAGAT presentation time are shown in Table 9.

**Table 9**

*Significance Tests for the Proportion of Correct Responses for the Scenario Identification Questions*

| Scenario Name | Haptics (*p*-values) | Question Presentation Time (*p*-values) |
| --- | --- | --- |
| Accident | **< .001** | .034 |
| Traffic Jam | **.017** | > .99 |
| Road Construction | .17 | .17 |
| Animal on Road | **.003** | .092 |
| Road Closed | **< .001** | .72 |

*Note. p*-values significant after FDR adjustment are presented in bold.

## Phase 2 Look Count, Game, and Satisfaction Results

**Look Count**

Boxplots depicting Look Count as the dependent variable are presented in Figure 17 and Figure 19. An outlier check using the 3xIQR criterion did not identify any outliers in the data.

**Figure 17**

*Look Count for the Presence of Haptic Cues*

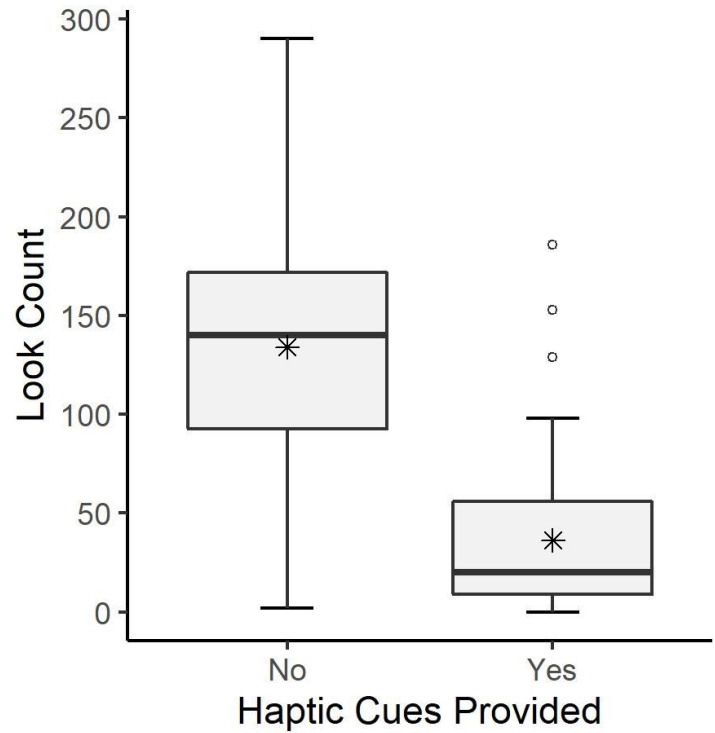

There was a significant main effect of haptic cuing $F(1, 130) = 114.37$, $MSE = 2796.72$, $p < .001$, and Cohen's $f = 0.94$. The conditions with haptic cues provided had lower Look Counts ($M = 36.31$, $Md = 20.00$) ranging from 0 to 186 compared to the conditions where cues were not provided ($M = 133.96$, $Md = 140.00$) ranging from 2 to 290. The main effect of SAGAT presentation time was not significant $F(1, 130) = 0.37$, $MSE = 2796.72$, $p = .55$, and Cohen's $f = 0.05$ with SAGAT during the drive conditions ($M = 83.09$, $Md = 67.00$) ranging from 0 to 290, and SAGAT after the drive conditions ($M = 87.18$, $Md = 66.00$) ranging from 0 to 253. The interaction between the presence of haptic cues and SAGAT presentation time was significant $F(1, 130) = 5.86$, $MSE = 2796.72$, $p = .017$, and Cohen's $f = 0.21$ (see Figure 18).

**Figure 18**

*Haptic Cuing and SAGAT Presentation Time Interaction for Look Count. Error bars represent a 95% Confidence Interval of the Mean.*

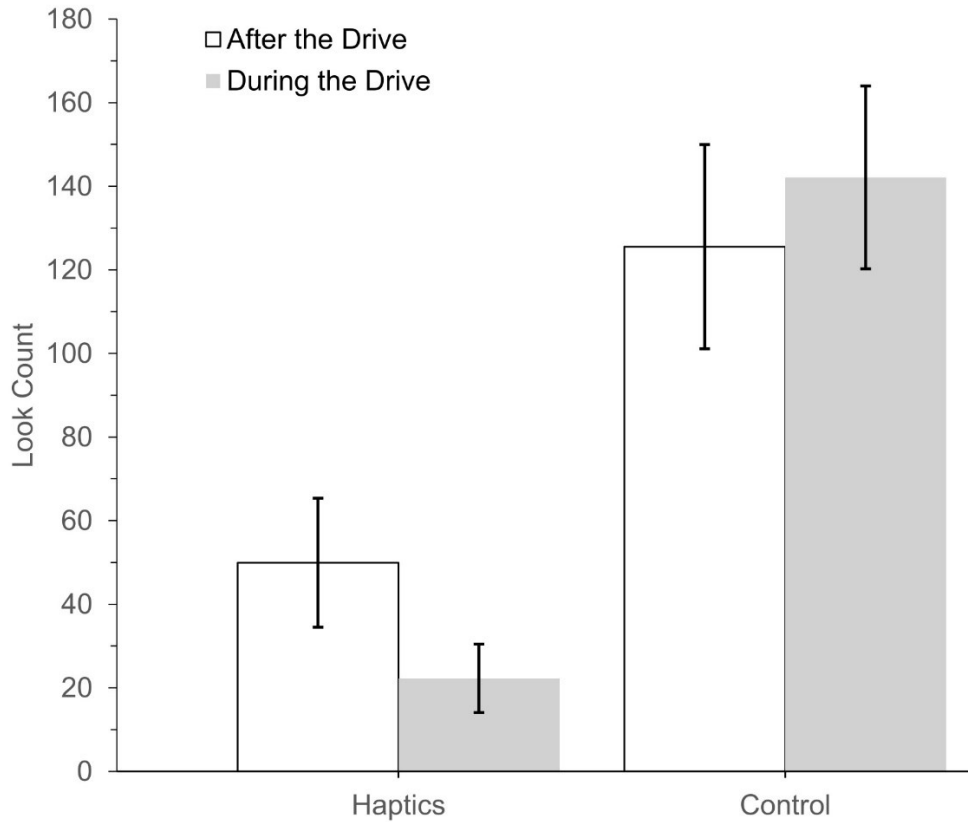

The interaction between the presence of haptic cues and SAGAT presentation time was decomposed further by comparing the effect of the presence of haptic cues on SAGAT presentation time. There was a statistically significant difference for Look Count for the cue-provided conditions depending on SAGAT presentation time with the SAGAT during the drive condition ($M = 22.27$) having a lower mean compared to the condition with SAGAT at the end of the drive ($M = 49.94$) $t(65) = 3.19$, $p = .002$, and Cohen's $d = 0.78$. There was no statistically significant difference for Look Count for the no-cue conditions between the SAGAT during the drive condition ($M = 142.11$) and the SAGAT at the end of the drive condition ($M = 125.54$) $t(65) = 1.03$, $p = .31$, and Cohen's $d = 0.25$.

**Figure 19**
*Look Count for the Question Presentation Time*

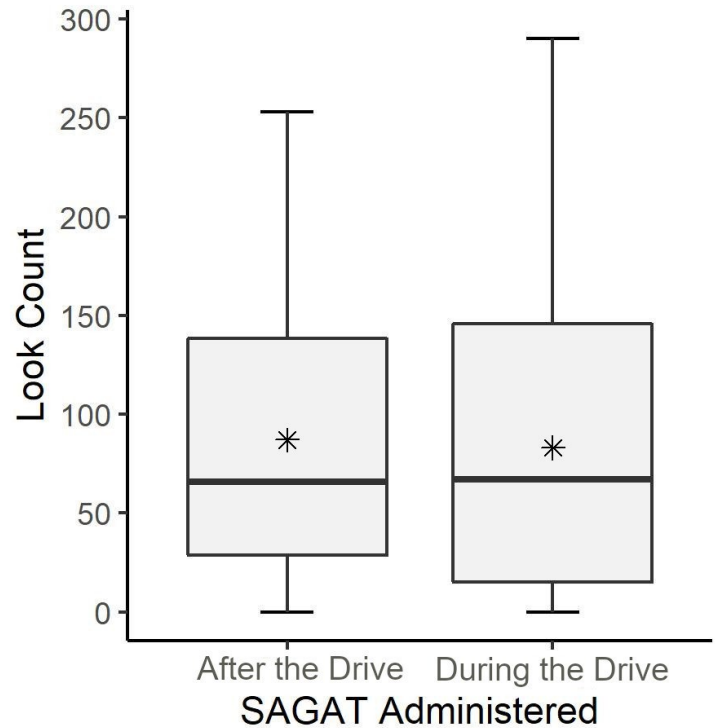

**Performance on Fruit Ninja**

Boxplots depicting these conditions with the number of games played as the dependent variable are presented in Figure 20 and Figure 21. An outlier check using the 3xIQR criterion did not identify any outliers in the data.

**Figure 20**

*Number of Games Played for the Presence of Haptic Cues*

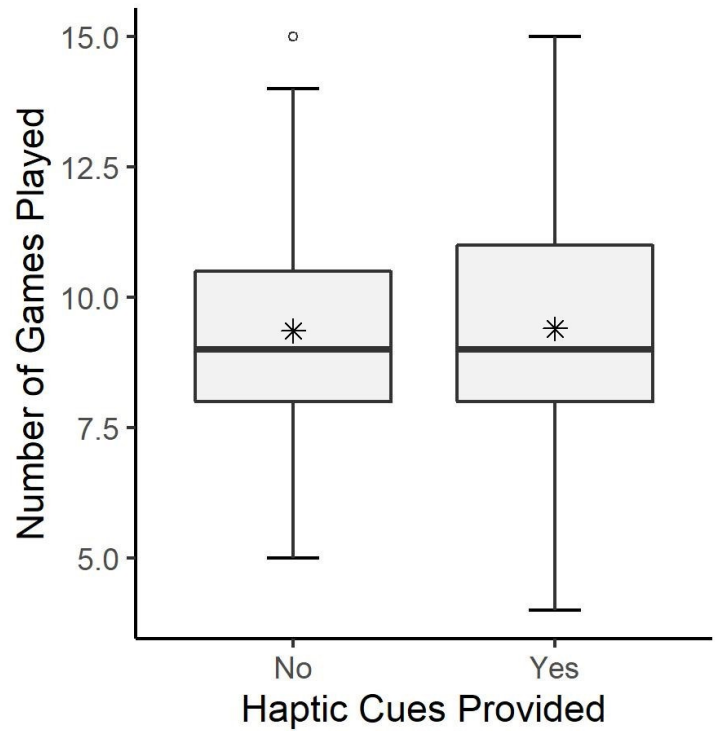

    The main effect of haptic cuing was not significant $F(1, 130) = 0.01$, $MSE = 4.48$, $p = .91$, and Cohen's $f = 0.01$ with cue-proved conditions ($M = 9.40$, $Md = 9.00$) ranging from 4 to 15, and no-cue conditions ($M = 9.36$, $Md = 9.00$) ranging from 5 to 15. The main effect of SAGAT presentation time was not significant $F(1, 130) = 0.28$, $MSE = 4.48$, $p = .60$, and Cohen's $f = 0.05$ with SAGAT during the drive conditions ($M = 9.28$, $Md = 9.00$) ranging from 5 to 15, and SAGAT after the drive conditions ($M = 9.48$, $Md = 9.00$) ranging from 4 to 15. The interaction between the presence of haptic cues and the SAGAT presentation time was also not significant $F(1, 130) = 1.68$, $MSE = 4.48$, $p = .20$, and Cohen's $f = 0.11$. There was no evidence based on the number of games played to suggest that providing haptic cues affected the engagement in playing Fruit Ninja.

**Figure 21**

*Number of Games Played for the Question Presentation Time*

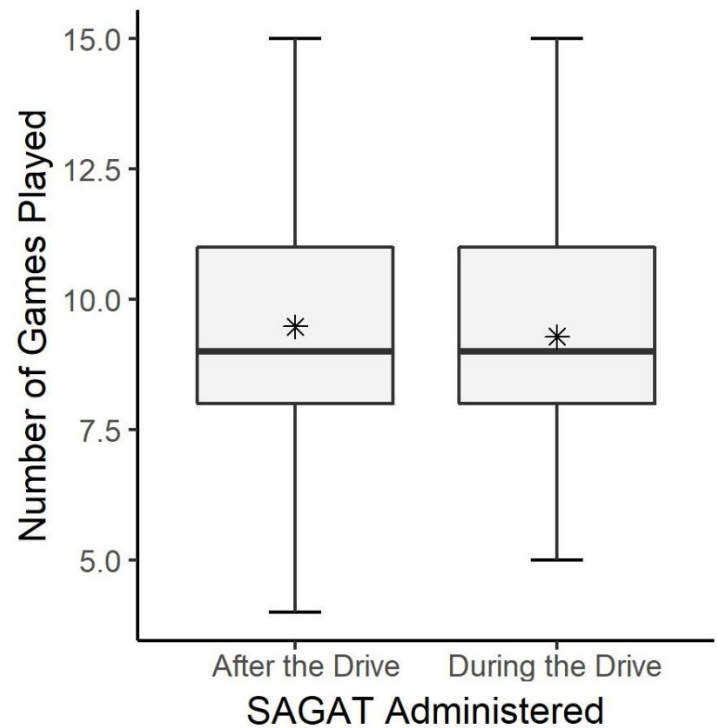

　　　　Boxplots comparing these conditions with the mean game score as the dependent variable are presented in Figure 22 and Figure 23. There was one 3xIQR outlier in the haptic cues provided conditions, but that outlier was not removed during analysis.

**Figure 22**

*Mean Game Score for the Presence of Haptic Cues. One 3xIQR outlier shown as a solid circle.*

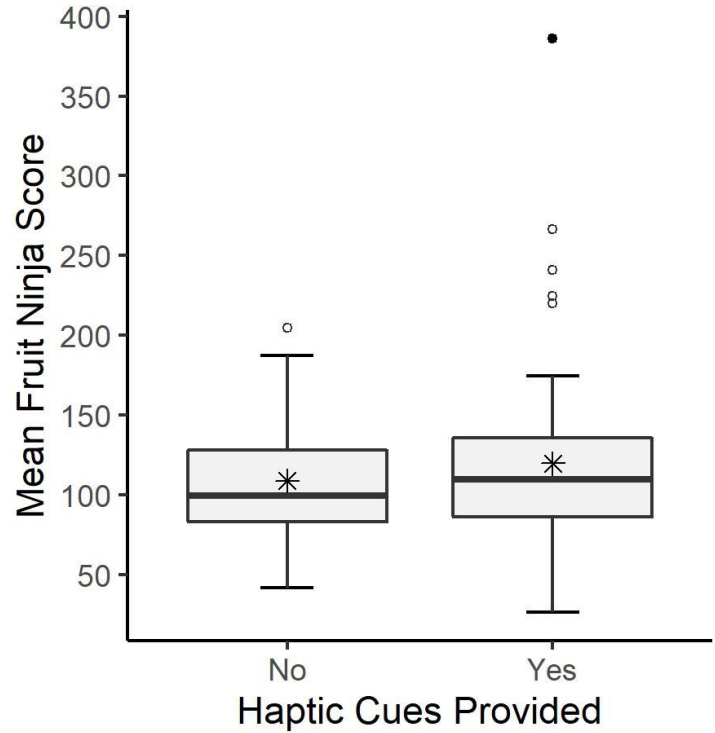

      The main effect of haptic cuing was not significant $F(1, 130) = 1.83$, $MSE = 2149.83$, $p = .18$, and Cohen's $f = 0.12$ with the cue-provided conditions ($M = 119.95$, $Md = 109.$) ranging from 26.87 to 385.75, and the no-cue conditions ($M = 108.97$, $Md = 99.81$) ranging from 41.73 to 204.67. The main effect of SAGAT presentation time was not significant $F(1, 130) = 1.84$, $MSE = 2149.83$, $p = .18$, and Cohen's $f = 0.12$ with SAGAT during the drive conditions ($M = 108.95$, $Md = 100.00$) ranging from 26.87 to 240.86, and SAGAT after the drive conditions ($M = 119.97$, $Md = 109.63$) ranging from 47.73 to 385.75. The interaction between the presence of haptic cues and the SAGAT presentation time was also not significant $F(1, 130) = 0.14$, $MSE = 2149.83$, $p = .71$, and Cohen's $f = 0.03$. There was no evidence based on game scores to suggest that providing haptic cues affected the engagement in playing Fruit Ninja.

**Figure 23**

*Mean Game Score for the Question Presentation Time. One 3xIQR outlier shown as a solid circle.*

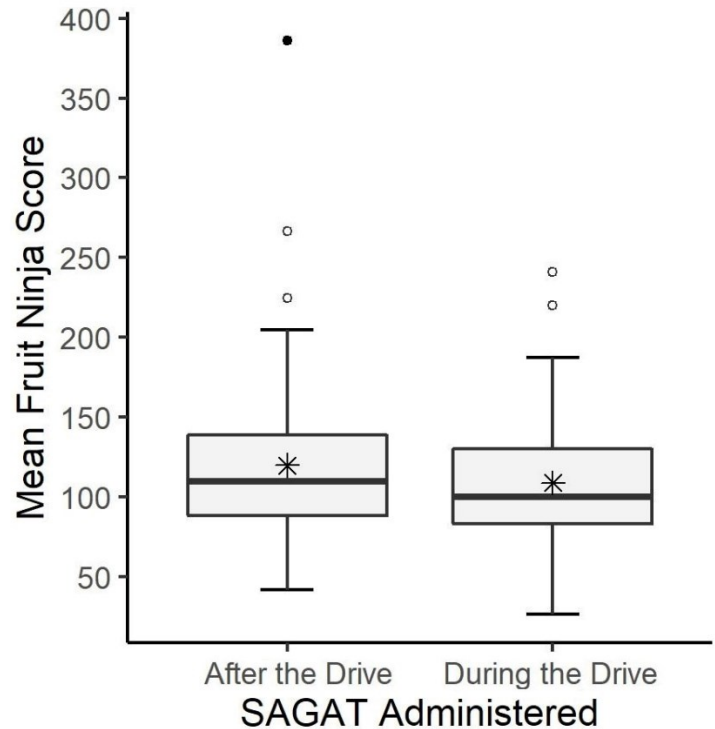

**Satisfaction**

The SUS scores from 5 subjects were discarded because they rated the haptic cues instead of the haptic device. There was no significant difference in SUS scores for the haptic device between condition HD ($M = 72.42$, $Md = 77.50$) ranging from 32.50 to 92.50 and condition HA ($M = 72.42$, $Md = 77.50$) ranging from 37.50 to 87.50, $t(61) = 0.96$, $p = .34$, and Cohen's $d = 0.24$ (see Figure 24). The adjective rating for SUS scores in both conditions is "good".

**Figure 24**
*SUS Scores for the Haptic Device for Conditions HD and HA*

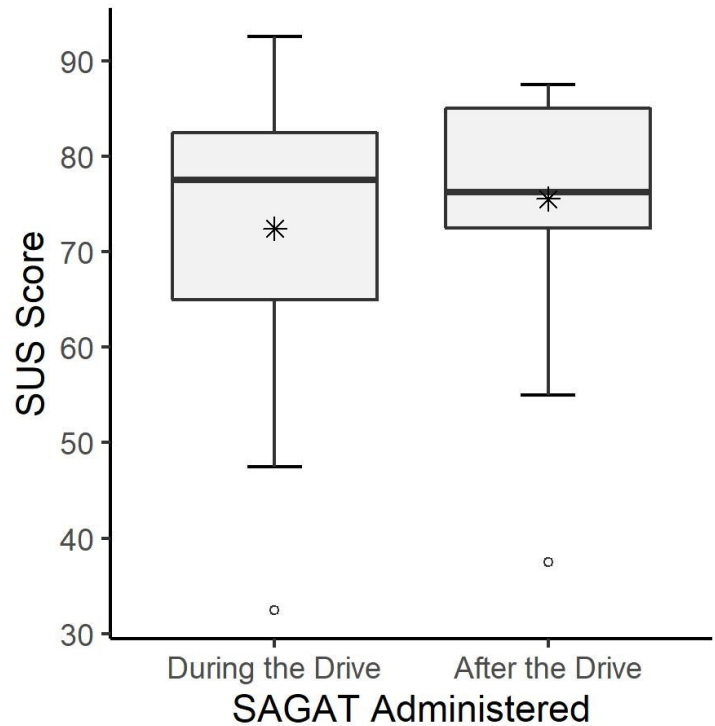

Additionally, in response to the statement whether the subjects found the cues to be disruptive rated on a scale of 1 to 5, with 1 being strongly disagree and 5 being strongly agree, there was no significant difference in the perceived disruptiveness of the cues for the haptic device between condition HD ($M = 2.06$, $Md = 2.00$) ranging from 1 to 4 and condition HA ($M = 1.97$, $Md = 2.00$) ranging from 1 to 5 (see Figure 25) $t(66) = 0.39$, $p = .70$, and Cohen's $d = 0.10$. This rating means subjects did not consider the haptic cues to be disruptive.

**Figure 25**

*Perceived Disruptiveness of Haptic Cues (lower is better) for Conditions HD and HA. Subjects were asked to rate the statement "I felt the vibration feedback was disruptive" on a scale of 1 to 5 with 1 being strongly disagree and 5 being strongly agree.*

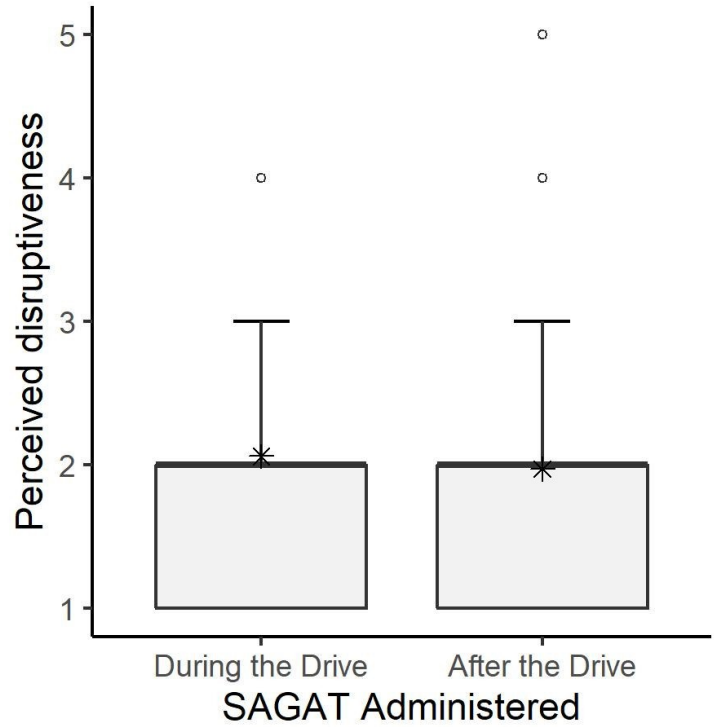

# Chapter 5
# Discussion

Can drivers' situation awareness during automated driving be maintained using haptic cues which provide information about road and traffic scenarios while the drivers are engaged in a secondary task? And can this be done without disengaging them from the secondary task? Subjects who received haptic cues had a higher number of correct responses to the situation awareness questions and looked up at the simulator screen fewer times than those who were not provided cues. Subjects did not find the cues to be disruptive and gave good satisfaction scores to the haptic device. Additionally, subjects across all conditions seemed to have performed equally well in playing Fruit Ninja. It does indeed appear that haptic cuing can maintain drivers' situation awareness during automated driving while drivers are engaged in a secondary task.

**Number of Scenarios Identified Correctly and Correct Responses per Scenario**

The main effects of haptic cuing for the dependent variables measuring the number of scenarios identified correctly and the number of correct responses per scenario were significant. This significance indicates that subjects were not only able to identify the haptic cues but also correctly associate them with the scenarios to identify the scenario at hand. It also indicates that drivers were able to use that information to correctly respond to the situation awareness questions presented to them. The effect of SAGAT presentation time was not significant for either variable indicating that subjects successfully recalled the information provided to them by the haptic cues for at least 20 minutes after they received the haptic cues, and that they may also successfully recall the order they received they received the cues in—further research would be needed to confirm whether that is indeed the case. The interaction between the presence of haptic cues and SAGAT presentation time was also not significant for either variable. Therefore, there was no evidence to suggest that the effect of haptic cues for different scenarios depended on whether the situation awareness questions are presented during the drive or at the end of the drive.

The within-subjects part of the analysis for the number of correct responses per scenario variable identified a significant effect of scenario type and a significant interaction between scenario type and haptic cuing. These findings indicated that one of the scenarios had fewer correct responses in the cue-provided conditions compared to the other scenarios. It was likely that cue design issues with Cue 3 might be the underlying cause for this difference. Indeed, an interaction contrast comparing the number of correct responses in the Road Construction scenario, which was associated with Cue 3, with the other four scenarios was significant indicating that it was likely that the fewer correct responses in the Road Construction scenario compared to the other scenarios were due to the cue design issues associated with Cue 3. This finding further highlights the fact that subjects were indeed relying on haptic cues to provide them the information they needed to respond to the questions associated with each scenario and their performance suffered when the cue could not be accurately and consistently distinguished from another cue. Thus, providing any random set of haptic cues in and of itself is not sufficient to update the situation awareness of drivers engaged in secondary tasks during automated driving, and care must be taken when designing the haptic cues to make sure that the cues are distinct enough to be recognized consistently and accurately.

**SAGAT Responses and Scenario Identification**

Overall, there was a significant advantage of haptic cuing in the following scenarios and levels of situation awareness:

1. The Accident Ahead scenario and the Traffic Jam for situation awareness Level 1.
2. The Traffic Jam and the Animal on Road scenarios for situation awareness Level 2.
3. The Road Closed scenario for situation awareness Level 3.
4. All scenarios except the Road Construction scenario, which was associated with the problematic Cue 3, were significant for haptic cuing for the scenario identification question.

Compared to haptic cuing, the effect of SAGAT presentation time was significant for fewer scenarios and levels of situation awareness:

1. The Accident Ahead and the Road Closed scenarios for situation awareness Level 3.
2. Only the Accident Ahead scenario for the scenario identification question.

It is likely that haptic cuing performed worse for individual situation awareness queries as compared to scenario identification questions because each individual SAGAT query was presented about 30 to 80 times over the course of the entire experiment which may have led to a loss in statistical power. Endsley recommends that each SAGAT query be presented at least 30 times over an entire experiment (Endsley, 2000). However, presenting some queries around 30 times over an entire experiment led to some experiment conditions having fewer than 10 individual SAGAT query responses per condition. In contrast, a scenario identification question was included for each scenario and for each subject thus ensuring that each scenario identification question was administered more than 128 times during the entire experiment. It is likely that this loss in statistical power may explain why haptic cuing performed worse for individual SAGAT queries when compared to scenario identification questions.

The data provide evidence that haptic cuing was effective for some scenarios for the levels of situation awareness, and that haptic cuing was effective for all but one scenario for scenario identification purposes. Therefore, subjects were able to use the haptic cues to identify the upcoming road scenarios and use the cues to respond correctly to the situation awareness questions while playing Fruit Ninja during autonomous driving.

**Look Count**

The presence of haptic cues had a large effect on the number of times subjects looked up at the simulator screen. The main effect of SAGAT presentation time was not significant. This indicates that subjects who were provided haptic cues relied on the cues to give them information about the upcoming road scenarios and as a result looked up at the simulator screen fewer times compared to those subjects who were not provided haptic cues. The statistically significant interaction of haptic cuing and SAGAT presentation time was because in the cues-provided conditions, subjects who responded to SAGAT questions during the drive looked up at the simulator screen fewer times compared to the subjects who responded to SAGAT questions at the end of the drive. One possible explanation for this is that subjects who responded to questions at the end of the drive likely looked up at the screen more to aid with retention. Additionally, this effect was not significant for subjects who were not provided haptic cues. This is likely because in the absence of haptic cuing subjects could only gather information about the upcoming scenarios by looking up at the simulator screen whenever they felt like it—thus there is no significant pattern in the no-cue conditions.

**Fruit Ninja Performance**

Neither main effects nor the interaction of haptic cuing and SAGAT presentation time were significant for the number of games played dependent variable. Therefore, there was no evidence to suggest that haptic cuing had an affect on task engagement in playing Fruit Ninja. Similarly, the main effects of haptic cuing and SAGAT presentation time were not significant for the mean game score variable. The interaction between the presence of haptic cues and SAGAT presentation time was also not significant. Therefore, based on mean game scores, there was also no evidence to suggest that haptic cuing had an affect engagement in playing Fruit Ninja.

Some subjects swapped out different swords and dojos which give different bonuses without changing the rules of the game in Classic Mode. This behavior was not limited to any specific condition in the experiment and indicated a natural level of engagement in the secondary task since some players are likely to try out different swords and dojos to find the one they like best when playing the game on their own. Amusingly, some subjects mentioned later that after participating in the experiment they picked up playing Fruit Ninja regularly again after many years.

**Haptic Device Satisfaction**

Subjects did not find the haptic cues to be disruptive and gave the haptic device satisfaction scores that would correspond to an adjective rating of "good". It was critical that subjects not find the cues to be disruptive and give good satisfaction ratings to the device otherwise one could reasonably conclude that drivers would be unlikely to use such a system. Since there were no statistically significant differences for SUS scores or perceived disruptiveness of the cues in the cue-provided conditions, thus one can conclude that there was no evidence to suggest that subjects did not find the device to be equally usable regardless of the SAGAT presentation time. Some users left comments mentioning that they confused Cue 3 for Cue 4 (originally, Cue 7 in Phase 1), and comments providing ideas for future iterations of the haptic device.

**Theoretical Implications**

Autonomous driving systems promise to free the driver from the task of actively driving the vehicle to varying degrees depending on their level of autonomy. This creates problems with loss of situation awareness in the driver since the driver is no longer actively involved in driving and may instead be engaged in NDRTs. The loss of situation awareness in the driver becomes an issue when the autonomous system encounters an error or a system boundary and issues a TOR for the driver to take over control, because drivers with low situation awareness may not be able to provide an appropriate and timely response to a TOR.

Prior research in using HMI in autonomous cars to address challenges due to the loss of situation awareness in drivers have focused on multiple levels of automation, on unimodal and multimodal TORs, on what happens once a TOR has been issued, and on using haptic cues for warning, informational, and guidance purposes. This experiment involved using haptic cues to maintain the situation awareness of drivers during Level 4 automated driving while the drivers were engaged in a secondary task and without disengaging them from the secondary task.

Driving engages our audio-visual senses, and drivers may pick up audio-visual NDRTs during highly autonomous driving. Multiple resources theory predicts that any information provided to the drivers using audio-visual modalities while they are engaged in audio-visual NDRTs is likely to suffer interference. Additionally, haptic cuing uses a sensory channel separate from our audio-visual senses and hence multiple resource theory predicts that any haptic stimuli are unlikely to suffer interference from audio-visual NDRTs. Therefore, haptic cuing provides a way to convey information to drivers engaged in audio-visual NDRTs while keeping the perceived disruptiveness at a low level.

      Indeed, consistent with predictions from multiple resource theory, subjects who received haptic cues had a higher number of correct responses to the situation awareness questions and looked up at the simulator screen fewer times than those who were not provided cues. Subjects did not find the cues to be disruptive and gave good satisfaction scores to the haptic device. Additionally, subjects across all conditions seemed to have performed equally well in playing Fruit Ninja and there was no evidence to suggest that the haptic cues interfered with the audio-visual NDRT of playing Fruit Ninja. Although these results are consistent with the predictions from multiple resource theory, they are not a strong test of multiple resource theory. A strong test of multiple resource theory would have included at least one modality that would have shown interference with the NDRT of playing Fruit Ninja.

      Haptic cuing used in this research was consistent with prior research in haptic driver support which includes issuing unimodal and multimodal TORs, issuing TORs to drivers while they were engaged in secondary tasks with the autonomous system driving the car, assisting the drivers in controlling the vehicle while they were driving, displaying spatial information about the traffic close to the drivers while the autonomous system was driving the car, and displaying warning messages and informational cues to subjects engaged in visual tracking tasks as substitutes for driving. In this experiment, subjects were provided haptic cues on the upper arm area of the arm they were not using to play Fruit Ninja, which is also consistent with prior research where studies have provided haptic feedback for driver support on seats, beneath the seats, seatbelts, wrists, upper arms, waists, steering wheels, gas pedals, and joysticks. Practical implications of the findings from this research and considerations based on comments from subjects regarding future iterations of the device are discussed in the next section.

**Practical Implications**

      Haptic cuing can maintain drivers' situation awareness during automated driving while drivers are engaged in a secondary task. Subjects who received haptic cues performed just as well on variables measuring secondary task engagement as those who did not receive the cues. Also, subjects did not find the cues to be disruptive and gave good satisfaction scores to the haptic device. There were some cue design issues with Cue 3 and some subjects mentioned ideas for future iterations of the device.

      Care must be taken while designing haptic cues since the drivers do rely on the information provided by the cues to obtain information about upcoming road conditions. Drivers are also likely to confuse the cues with one another or not recall their association to the corresponding road conditions if the cues are not distinguishable from each other on an accurate and consistent basis.

      Subjects seemed to remember the cue associations and the order cues were displayed in for at least 20 minutes in this research. Since Level 4 autonomous systems are quite capable of driving under a wide range of road conditions, it may be worthwhile in practice to implement a lower frequency for haptic cuing, perhaps a few times per hour depending on the scenario at hand. It may also be helpful to allow the drivers to select some additional scenarios for haptic cuing beyond a default set implemented in the autonomous system.

      The haptic device used in this research was a prototype proof of concept and should not be implemented as-is in vehicles. Future research should evaluate the comparative satisfaction of providing these cues using seatbelts, seat backs, or the seat bottom for instance. Some subjects

told the experimenters that they liked the idea of the device but would prefer if it did not have to be worn separately. Providing these cues using any of three areas mentioned previously would address this concern. Another subject commented about implementing the cues as bursts of air; however, it is possible that any such implementation may be confused with the vehicle's own air conditioning and heating system.

Haptic cuing provides a countermeasure for the loss of situation awareness during automated driving while drivers are engaged in secondary tasks and without having an impact on the drivers' engagement in audio-visual secondary tasks. Care must be taken while designing haptic cues to ensure they can be distinguished accurately and consistently. Additionally, further research is needed to identify the best location in the vehicle to display these cues.

**Limitations**

There were several limitations in this research. First, individual SAGAT queries were presented about 30 to 80 times over the course of the entire experiment which may have led to a loss in statistical power. Endsley recommends that each SAGAT query be presented at least 30 times over an entire experiment (Endsley, 2000). However, presenting some queries around 30 times over an entire experiment led to some experiment conditions having fewer than 10 individual SAGAT query responses per condition. It would have been ideal to have at least 30 query responses for each individual query per experiment condition, likely by reducing the number of SAGAT queries for each level of situation awareness. This reduction would have increased the total number of times each query was presented in the experiment thus addressing the issue of low statistical power for individual SAGAT queries.

Second, the subject pool for this experiment was limited to undergraduates at Rice University. Recruiting a more diverse subject pool was not feasible due to COVID-19 restrictions and time limitations. These restrictions limited the driving experience in the subject pool. A more diverse subject pool, comprised of subjects from different age groups and education levels, may have uncovered additional findings regarding the effectiveness of haptic cuing of road scenarios and the haptic device satisfaction ratings.

Third, there were no TORs implemented in the experiment because I did not have access to a Level 4 autonomous driving capable driving simulator. Level 4 autonomous driving was simulated by using road and traffic scenario screen recordings obtained while I drove through the scenarios in STISim Drive driving simulator. While this workaround allowed me to collect the data needed for my dissertation, I missed out on the opportunity to gain further insights by comparing driver takeover performance in the cue-provided vs. no-cue conditions.

**Future Directions**

There are many opportunities to for future research in driver interaction with Level 4 autonomous systems. In this experiment, subjects successfully recalled the information provided to them by the haptic cues for at least 20 minutes after they received the haptic cues, and seemed to also successfully recall the order they received they received the cues in. Future research could explore this further to determine whether there is a time limit on how long the information provided by haptic cues is available to drivers while they are engaged in secondary tasks and a Level 4 autonomous system is driving. Future research could also confirm whether drivers can indeed remember the order of haptic cues for an extended period, and the time limit for recalling that information.

It would also be worthwhile to research the frequency of haptic cuing of road conditions in Level 4 autonomous cars to figure out a good balance between alert fatigue and providing the necessary information to the drivers at reasonable intervals. The device used in this research was a prototype and future research could conduct a comparative evaluation of driver satisfaction with providing haptic cues associated with road scenarios using seatbelts, seat backs, or the seat bottom.

    There is a tremendous opportunity to conduct similar research recruiting drivers with disabilities. Higher levels of autonomy may make cars accessible to a wider range of drivers who cannot be served by the autonomous systems currently available in the market. Future research could help design HMI, including haptic cuing, in Level 4 autonomous vehicles to better serve the needs to drivers with disabilities.

    In this is research, there was no evidence to suggest that haptic cuing had a negative impact on the NDRT of playing Fruit Ninja. There was a concern that haptic cuing may impose an additional cognitive burden on the subjects as they recognized and associated each cue with its corresponding road scenario while playing Fruit Ninja. Since there was no evidence to suggest that haptic cuing has a negative impact on the NDRT, it is likely that recognizing the cues and associating them with the correct road scenario did not impose an additional cognitive burden on the drivers. However, future research could investigate additional haptic cue designs to identify any patterns that may potentially lower the cognitive burden on the subjects thereby freeing up some of their cognitive resources which in turn may allow haptic cuing to help improve subjects' performance on the NDRT compared to the NDRT performance in control conditions.

HAPTIC CUING OF ROAD CONDITIONS IN AUTONOMOUS CARS 57

HAPTIC CUING OF ROAD CONDITIONS IN AUTONOMOUS CARS    58

**Appendix A: Cue Visualizations and Python Scripts for Playing the Haptic Cues**

Cue visualizations depict the entire cue in the Syntacts GUI. The Play All file includes the code to cycle through all seven cues, and the Heartbeat file includes code to design the heartbeat haptic cue.

**Cue Visualizations**

**Cue 1: Jackhammer**

The cue called Jackhammer was a 200Hz Sine wave, duration = 1.5s, amplitude = 0.5, pulse-width modulated at 10Hz with a 0.3 duty cycle and scalar addition of 1 presented simultaneously on both tactors. The visualization shows the presentation on one tactor.

**Figure 26**
*Visualization of the Jackhammer cue.*

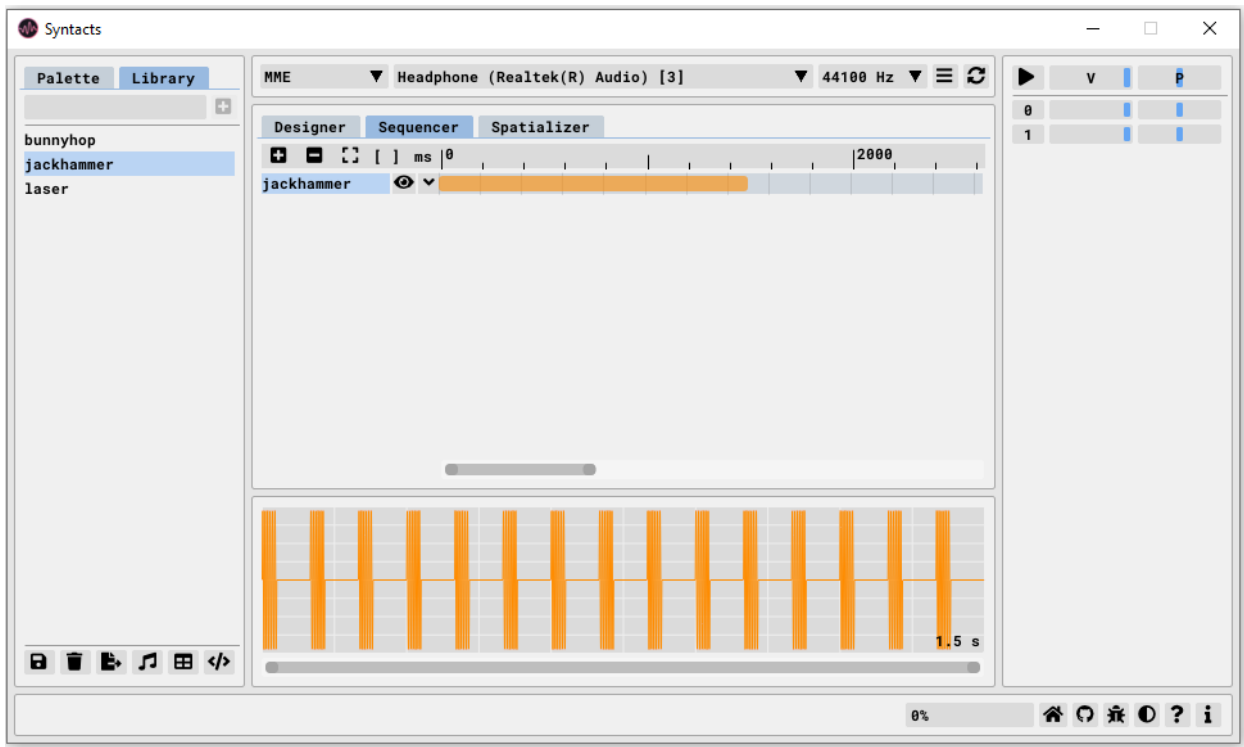

**Cue 2: Laser**

The cue called Laser was a 200Hz Square wave, duration = 0.5s, amplitude = 0.4 presented once on Tactor 1 then on Tactor 2 after a 0.05s wait.

**Figure 27**

*Visualization of the Laser cue. The wait time has been rounded up to 0.1 s since the Syntacts GUI is unable to show wait times smaller than 0.1s.*

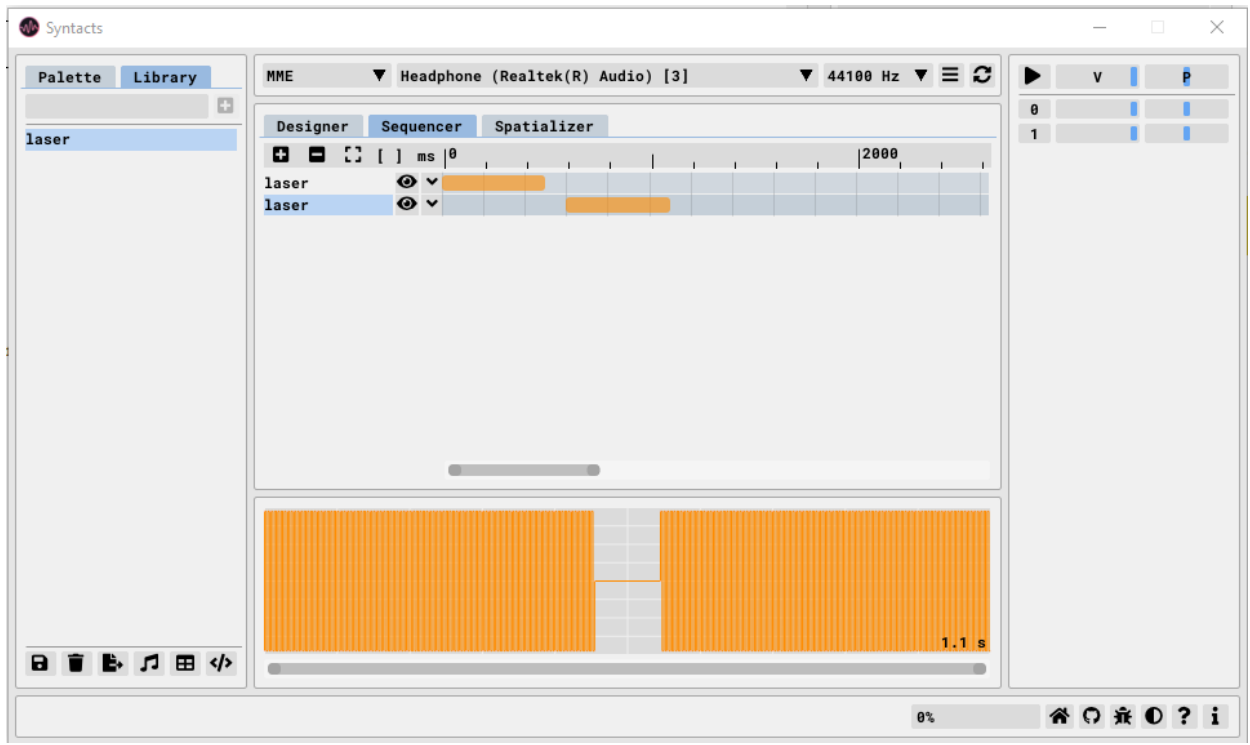

**Cue 3: Bunny Hop**

The cue called Bunny Hop was a 256Hz Square wave, duration = 0.05s, amplitude = 0.4 presented 4 times on Tactor 1 then on Tactor 2 after a 0.05s wait. The visualization shows the presentation on one tactor.

**Figure 28**

*Visualization of the Bunny Hop cue. The wait time has been rounded up to 0.1 s since the Syntacts GUI is unable to show wait times smaller than 0.1s.*

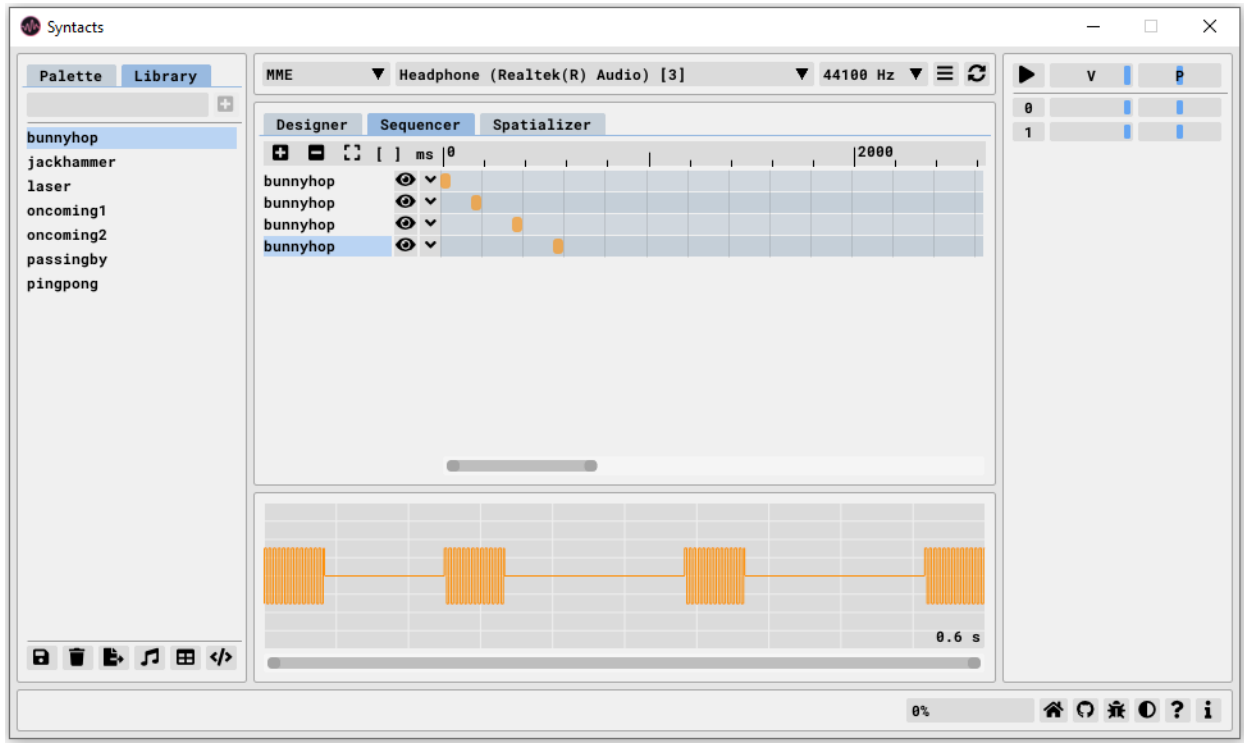

**Cue 4: Passing By**

The cue called Jackhammer was a 200Hz Sine wave, duration = 1.5s, amplitude = 0.5, pulse-width modulated at 10Hz with a 0.3 duty cycle and scalar addition of 1 presented simultaneously on both tactors. The visualization shows the presentation on one tactor.

**Figure 29**

*Visualization of the Passing By cue.*

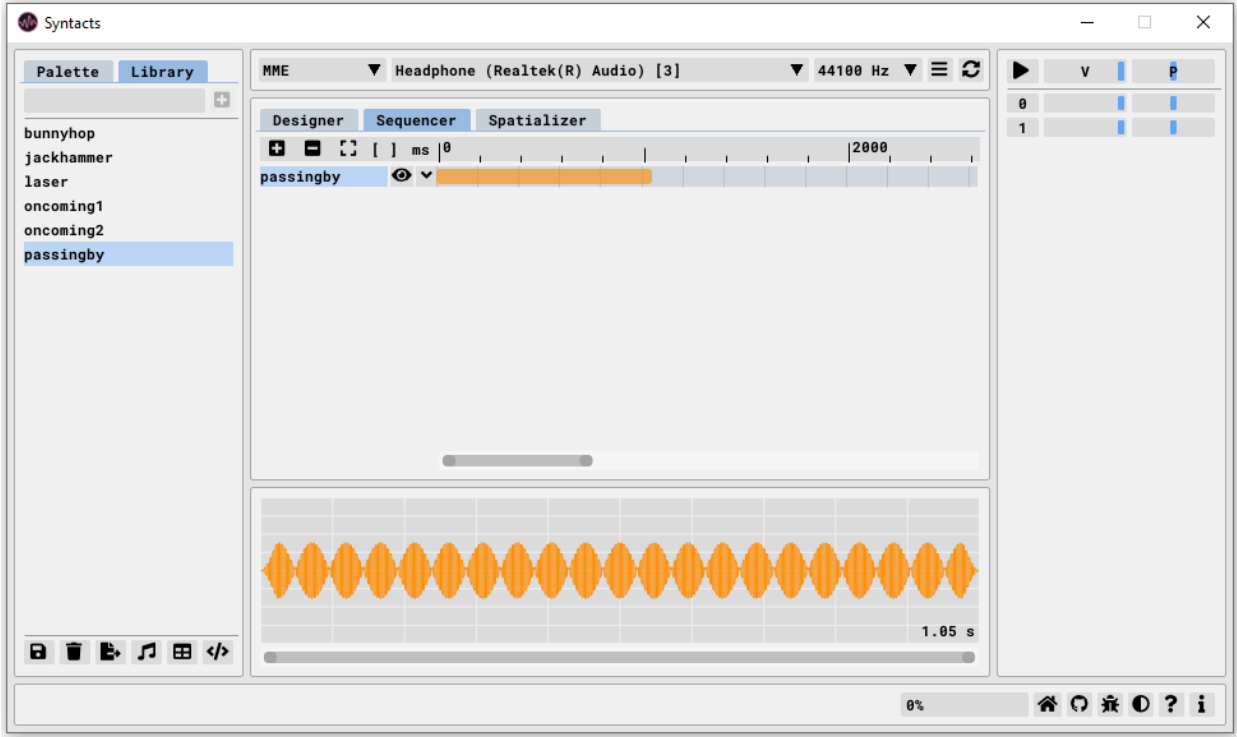

**Cue 5: Oncoming**

The cue called Oncoming was a two-part cue, with the first part as a 200Hz Square wave, duration = 1s, amplitude = 1 with a ramp modification starting at 0.01 and a rate of 1. Second part as a 300Hz Sine wave, duration = .1s, amplitude = 1. Presented as a combination with the first part displayed on Tactor 1 and the second part displayed on Tactor 2 with a 0.05s wait in between.

**Figure 30**
*Visualization of the Oncoming cue. The wait time has been rounded up to 0.1 s since the Syntacts GUI is unable to show wait times smaller than 0.1s.*

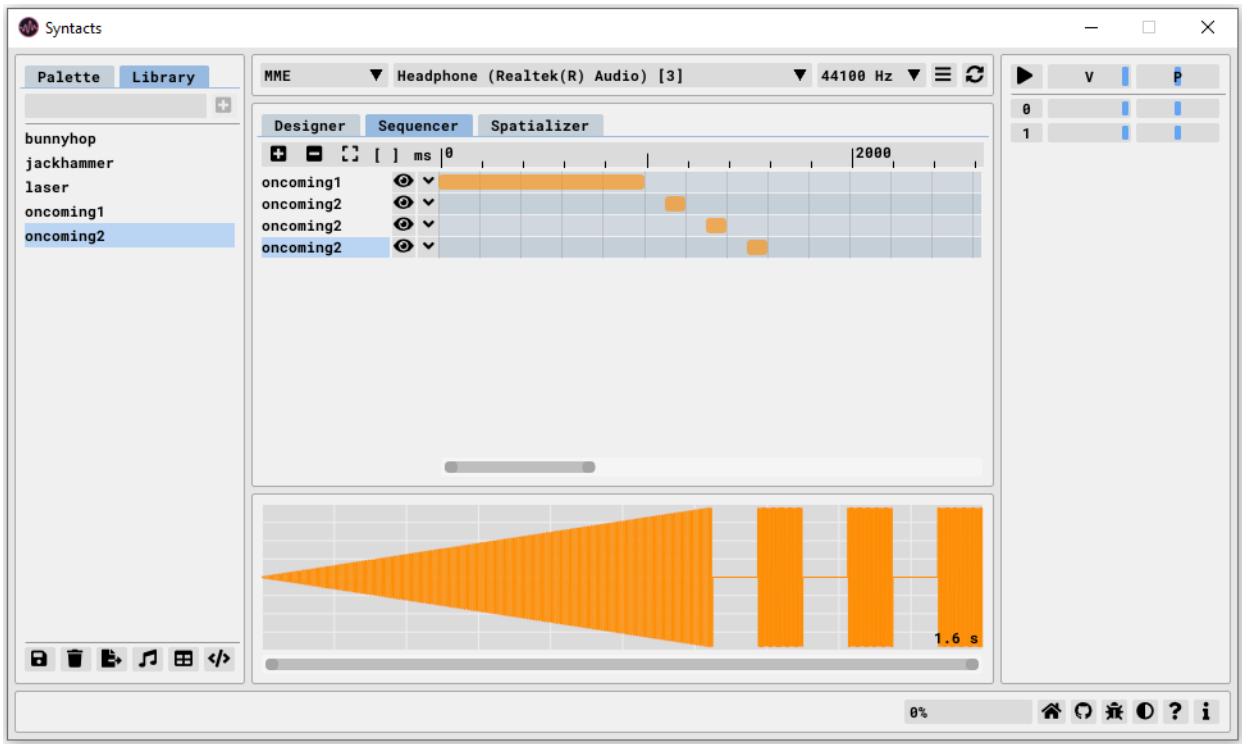

**Cue 6: Heartbeat**

The cue called Heartbeat was designed in two parts with the first half-beat as a 440Hz Sine wave Attack Sustain Release (ASR) of .1s .15s and .1s, and second half-beat at 0.75 times the first half-beat with a .1s gap in between the two combined to create one full heartbeat. Presented on repeat for 1.75s on each tactor with a .1s delay between the presentations. The visualization shows the presentation on one tactor.

**Figure 31**
*Visualization of the Heartbeat cue.*

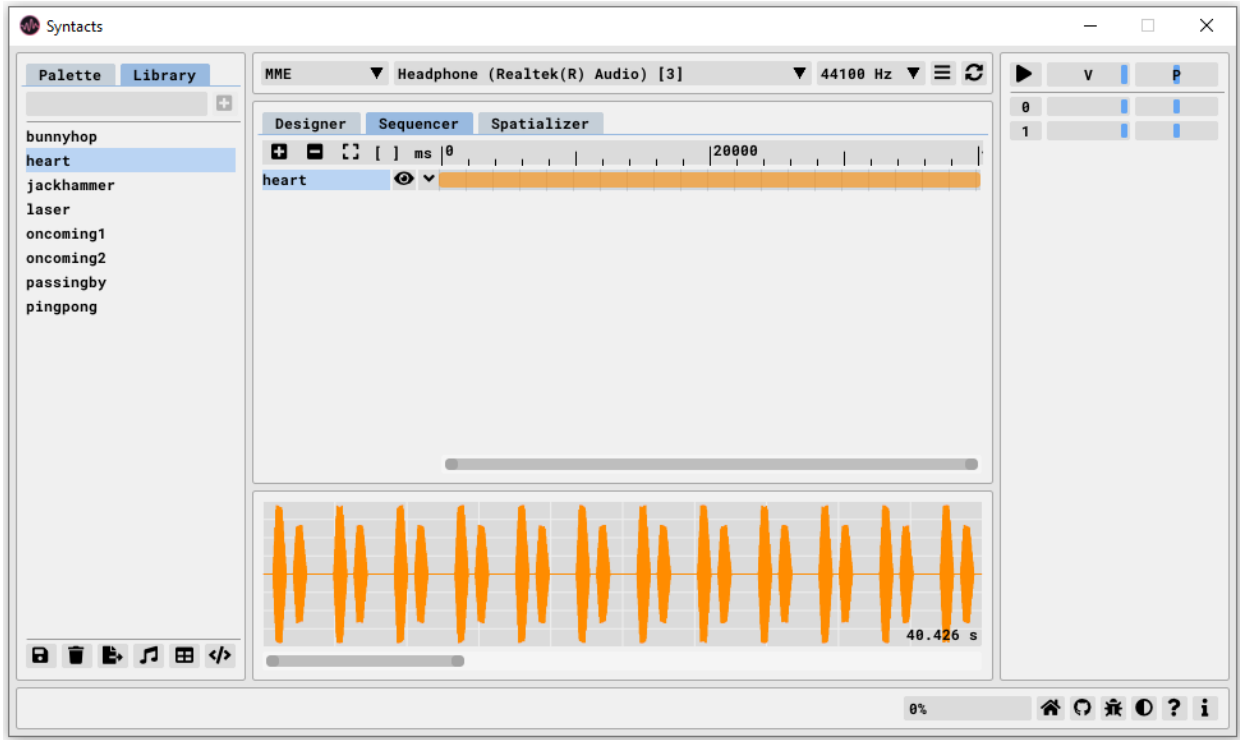

**Cue 7: Ping Pong**

The cue called Ping Pong was a 256Hz Square wave, duration = 0.05s, amplitude = 0.4 presented 5 times alternating between Tactor 1 and Tactor 2 with a 0.05s wait in between. The visualization shows alternating presentation on both tactors.

**Figure 32**

*Visualization of the Ping Pong cue. The wait time has been rounded up to 0.1 s since the Syntacts GUI is unable to show wait times smaller than 0.1s.*

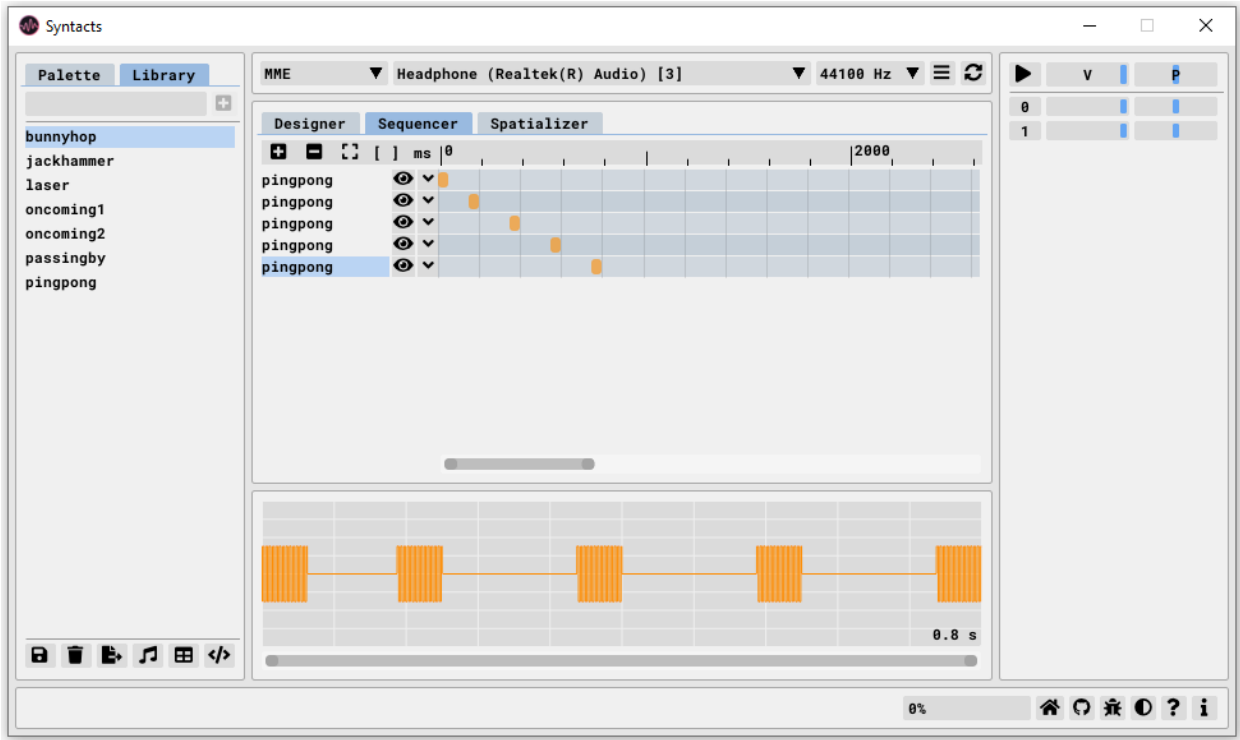

**Play All File**

```python
from syntacts import *
from time import sleep
from math import sin
from math import pi

#--------------------------------------------------------------
# Function to make sure export/import works
def check(signal):
    if signal is None:
        raise Exception("Import/Export failed.")

# ---------------------------------------------------------------------------

# Syntacts usage begins with creating an audio context, or Session
session = Session()
session.open()

# ---------------------------------------------------------------------------

# Vibrations are represented by Signals and combinations of Signals

sig1 = (Pwm(10, 0.3) + 1) * Envelope(1.5, 0.4) * Sine(200) # 10Hz PWM for 1.5 sec

sig2 = Square(200) * Envelope(.5, 0.4) # 200Hz square, duration is 0.5s, and amplitude is 0.4

sig3 = Square(256) * Envelope(.05, 0.4) # 256Hz square, duration is 0.05s, and amplitude is 0.4

sig4 = Pwm(230) * Sine(10) * ASR(.25, 1, .25, 0.4) # 230Hz PWM with 10Hz Sine and 1.5s duration ASR amplitude is 0.4

sig5 = Square(200) * Envelope(1) * Ramp(.01, 1) # first half of the cue
sig6 = Square(300) * Envelope(.1) # second half of the cue

# Import sig7
sig7 = Library.import_signal("heart.sig")
check(sig7)

#-----------------------------------------------------------------------

# Define separate functions to play each cue

def play_cue1():
    print("Playing cue 1: Jackhammer")
    # Play sig 1 once on all channels
    session.play_all(sig1)
    sleep(sig1.length)

def play_cue2(channel = 0):
    print("Playing cue 2: Laser")
```

```python
        # Play sig 2 wait 0.05s once on each channel
        session.play(0, sig2)
        sleep(sig2.length)
        sleep(0.05) # wait for 0.05s
        session.play(1, sig2)
        sleep(sig2.length)

def play_cue3():
    print("Playing cue 3: Bunny hop")
    # Play sig 3 wait 0.05s x 4 on channel 0 then play same on channel 1
    for x in range(4):
        session.play(0, sig3)
        sleep(sig3.length)
        sleep(.05)
    for x in range(4):
        session.play(1, sig3)
        sleep(sig3.length)
        sleep(.05)

def play_cue4():
    print("Playing cue 4: Passing by")
    # Play sig 4 once on channel 1
    session.play(1, sig4)
    sleep(sig4.length)

def play_cue5():
    print("Playing cue 5: Oncoming")
    # Play sig 5 on channel 0 wait 0.05s then play sig 6 thrice on channel 1
    session.play(0, sig5)
    sleep(sig5.length)
    sleep(.05)
    for x in range(3):
        session.play(1, sig6)
        sleep(sig6.length)
        sleep(.05)

def play_cue6():
    print("Playing cue 6: Heartbeat")
    # Play sig 7 once for 1.75 second on each channel with 0.1s delay in between
    for x in range(2):
        session.play(x, sig7)
        sleep(1.75)
        session.stop(x) # end cue playback
        sleep(0.1) # wait for 0.1s in between

def play_cue7():
    print("Playing cue 7: Ping pong")
    # Play sig 3 alternating between both channels 5 times per channel
    ch = [0, 1, 0, 1, 0, 1, 0, 1, 0, 1]
    for x in range(10):
        session.play(ch[x], sig3)
        sleep(sig3.length)
```

```python
        sleep(.05)

def all_cues():
    print("Playing all cues with a 2-second delay between them.")
    sleep(2)
    for x in range(1,8):
        eval("play_cue" + str(x) + "()") # Use eval to call functions dynamically
        sleep(2) # Wait 2 seconds

# Implement a custom switch-case for the functions

def switch_case(i):
    switcher = {
        1: play_cue1,
        2: play_cue2,
        3: play_cue3,
        4: play_cue4,
        5: play_cue5,
        6: play_cue6,
        7: play_cue7,
        8: all_cues
    }
    # choice will store the function from the switcher
    choice = switcher.get(i, "Please enter a number between 1 and 8.")
    # execute the function stored in choice to play the cue
    choice()

# Print out a welcome message and tell the user what this program does
print("\nWelcome to the Computer-Human Interaction Lab at Rice.\n")
print("Entering a number between 1 - 7 will play the corresponding cue.\n")
print("Entering 8 will play all cues with a 2-second delay in between them.\n")
print("Enter 9 to exit.")

# Now ask the user to pick which cue to play
cue_choice = 0
while cue_choice < 9:
    try:
        cue_choice = int(input('\nEnter a number between 1 and 9:'))
    except Exception:
        print("Please enter a valid number.")
        raise

    if cue_choice >= 1 and cue_choice <=8:
        switch_case(cue_choice) # look in the dictionary which cue to play
else:
    print("\nThank you for participating in our experiment.")

#------------------------------------------------------------------------

# Devices will automatically close when the Session goes out of scope,
# but it is good practice to do this explicitly
session.close()
```

**Heartbeat File**
```python
from syntacts import *
from time import sleep
# ----------------------------------------------------------------------

# Syntacts usage begins with creating an audio context, or Session
s = Session()
s.open()

# Create the cue
lub = Sine(440) * ASR(.1,.15,.1) # Sine with ASR
dub = lub * 0.75
lubdub = lub << 0.1 << dub # One beat

heartbeat = Repeater(lubdub, 3, 0.4) # Repeat to simulate a heartbeat

# Play it
s.play(0,heartbeat)
sleep(heartbeat.length)
#-----------------------------------------------------------------------

# Devices will automatically close when the Session goes out of scope,
# but it is good practice to do this explicitly
s.close()
```

**Appendix B: Driving simulator setup**

**Figure 33**

*Computer station used to play the drive video and display questions.*

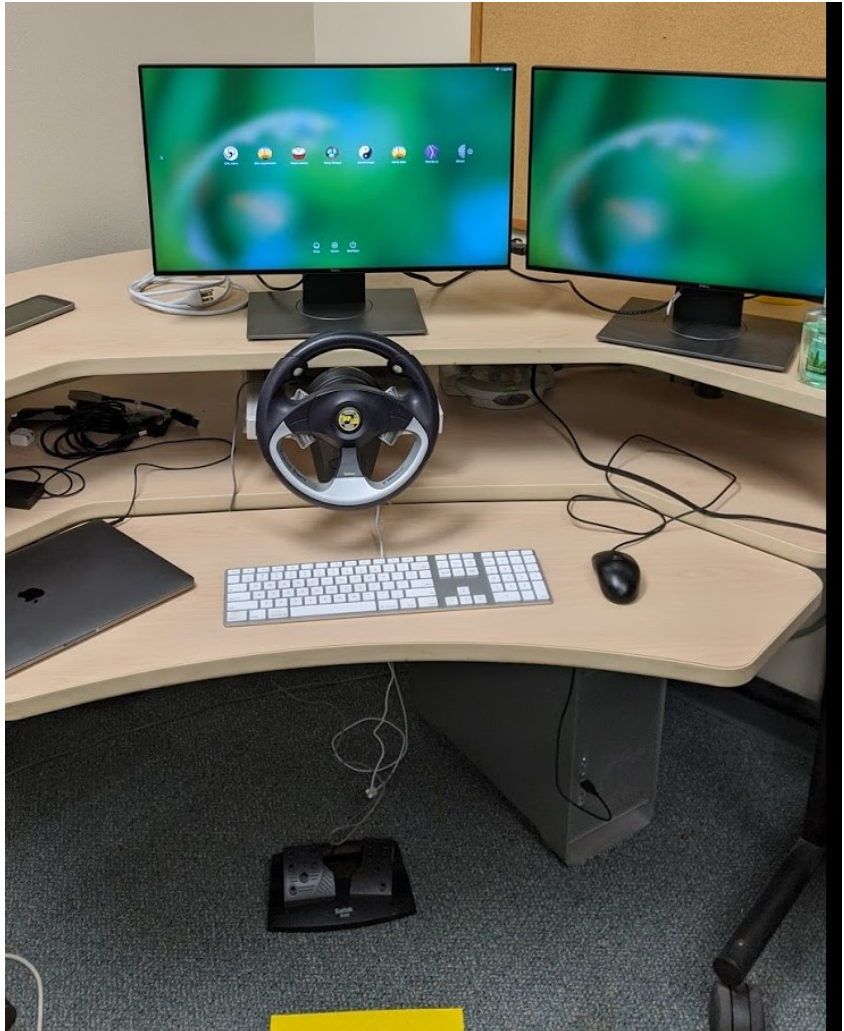

     Level 4 autonomous driving was simulated by showing a video on the left monitor and situation awareness questions were presented on the right monitor. To keep viewing distances similar to those in the STISim Drive Driving simulator, the monitors were arranged such that the viewing distance from the chair to the center of the left monitor was 25 inches from the eye of a person who is 5 foot 7 inches tall.

**Appendix C: Correct Responses per Scenario with Outliers Removed**

A 2x2x5 mixed design ANOVA was used to compare the experiment conditions using the number of correct responses per scenario for haptic cuing and for SAGAT presentation time. Bar charts comparing these conditions with number of correct responses per scenario as the dependent variable are presented in Figure 34 and Figure 35.

**Figure 34**

*Number of Correct Responses per Scenario for the Presence of Haptic Cues. Error bars represent +/- 1 Standard Error of the Mean.*

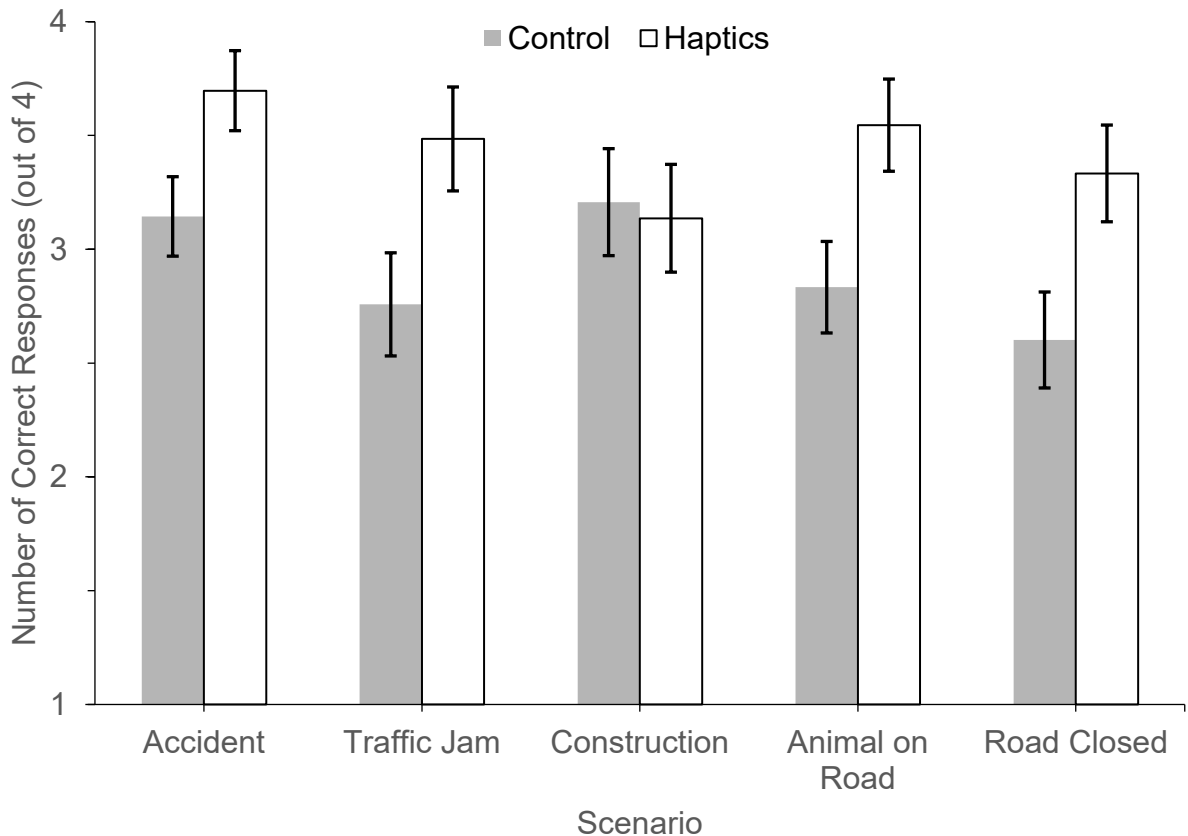

There were 7 within-subjects outliers across all five scenarios combined and one between-subjects outlier identified using the 3xIQR criterion. The between-subjects outliers were replaced with the overall subject mean calculated with the outliers removed and the between-subjects outlier was removed for the analysis.

**Figure 35**

*Number of Correct Responses per Scenario for Question Presentation Time. Error bars represent +/- 1 Standard Error of the Mean.*

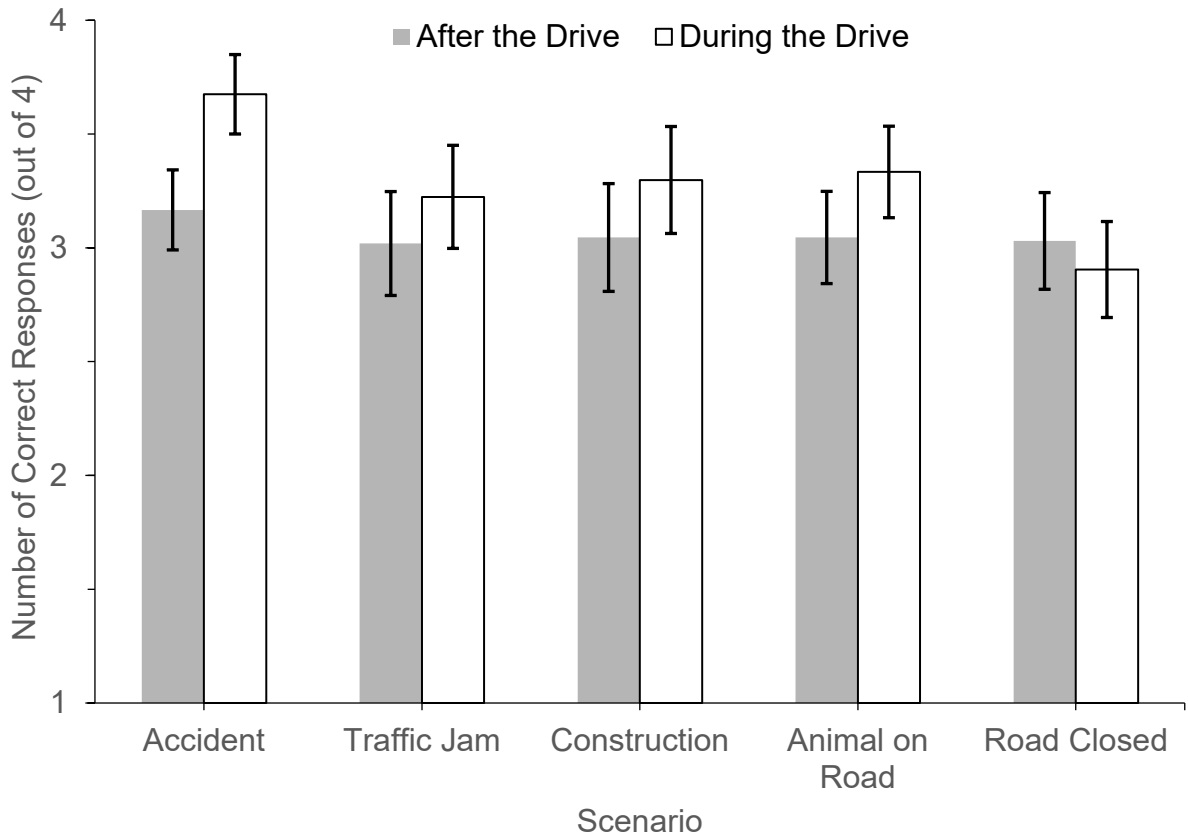

In the between-subjects part of the analysis, there was a significant main effect of haptic cuing $F(1, 129) = 15.15$, $MSE = 15.44$, $p < .001$, and Cohen's $f = 0.34$ comparing the cue-provided and the no-cue conditions. The main effect of SAGAT presentation time was not significant $F(1, 129) = 2.74$, $MSE = 15.44$, $p = .10$, and Cohen's $f = 0.15$ when comparing SAGAT after the drive conditions with SAGAT during the drive conditions. The interaction between the presence of haptic cues and the SAGAT presentation time was also not significant $F(1, 129) = 0.77$, $MSE = 15.44$, $p = .38$, and Cohen's $f = 0.08$.

In the within-subjects part of the analysis, sphericity was assumed for the main effect of scenario type and the interactions of scenario type with the between-subjects variables as the Huynh-Feldt test reported a value greater than .95. There was a significant effect of scenario type $F(4, 516) = 3.26$, $MSE = 1.09$, $p = .012$, and Cohen's $f = 0.16$. The interaction between scenario type and haptic cuing was also significant $F(4, 516) = 3.63$, $MSE = 1.09$, $p = .006$, and Cohen's $f = 0.17$. The interaction between scenario type and SAGAT presentation time was not significant $F(4, 516) = 1.59$, $MSE = 1.09$, $p = .18$, and Cohen's $f = 0.11$. The three-way interaction between scenario type, haptic cuing, and SAGAT presentation time was also not significant $F(4, 516) = 2.06$, $MSE = 1.09$, $p = .085$, and Cohen's $f = 0.13$.

The interaction between scenario type and haptic cuing was decomposed further with an interaction contrast comparing the Road Construction scenario with the other four scenarios. This contrast was chosen because the Road Construction scenario was associated with Cue 3 and also based on Phase 1 findings which indicated that subjects confused Cue 3 and Cue 7 with each other about 7.14% of the time and based on comments from some subjects in Phase 2 that they found the cue for the Road Construction scenario and the cue for the Animal on Road scenario (associated with Cue 7) to be similar. This interaction contrast would help uncover whether the number of correct responses were indeed fewer in the Road Construction scenario as compared to the other four scenarios.

The interaction contrast comparing the number of correct responses in the Road Construction scenario with the other four scenarios was significant $F(1, 129) = 11.36$, $MSE = 26.47$, $p = .001$, and Cohen's $f = 0.30$. This tells us that there were fewer number of correct responses in the Road Construction scenario as compared to the other four scenarios. It is likely that the fewer number of correct responses in the Road Construction scenario were due to the cue design issues associated with Cue 3.

## Appendix D: Significance tests for individual SAGAT Questions

FDR-adjusted significance tests for the proportion of correct responses using Fisher's Exact Test to compare haptic cuing and the time situation awareness questions were asked for all individual SAGAT queries associated with all three levels of situation awareness for all five scenarios are provided in this appendix.

**Responses to individual Level 1 SAGAT Questions**

**Table 10**

*Significance Tests for the Proportion of Correct Responses for Individual SAGAT Questions for Situation Awareness Level 1*

| Query Name | Haptics (p-values) | Question Presentation Time (p-values) |
| --- | --- | --- |
| Accident Query 1 | **< .001** | .091 |
| Accident Query 2 | .53 | .54 |
| Traffic Jam Query 1 | .11 | > .99 |
| Traffic Jam Query 2 | **.035** | .054 |
| Road Construction Query 1 | > .99 | .053 |
| Road Construction Query 2 | .59 | .79 |
| Animal on Road Query 1 | .01 | .61 |
| Animal on Road Query 2 | .11 | .31 |
| Animal on Road Query 3 | .73 | **< .001** |
| Road Closed Query 1 | .56 | .084 |
| Road Closed Query 2 | .57 | > .99 |

*Note.* p-values significant after FDR adjustment are presented in bold.

**Responses to individual Level 2 SAGAT Questions**

**Table 11**

*Significance Tests for the Proportion of Correct Responses for Individual SAGAT Questions for Situation Awareness Level 2*

| Query Name | Haptics (p-values) | Question Presentation Time (p-values) |
| --- | --- | --- |
| Accident Query 1 | > .99 | .26 |
| Accident Query 2 | .18 | **.041** |
| Accident Query 3 | .01 | .61 |
| Accident Query 4 | .60 | .61 |
| Traffic Jam Query 1 | .059 | .20 |
| Traffic Jam Query 2 | .25 | .56 |
| Traffic Jam Query 3 | > .99 | .27 |
| Road Construction Query 1 | .06 | .19 |
| Road Construction Query 2 | .48 | > .99 |
| Road Construction Query 3 | .30 | > .99 |
| Animal on Road Query 1 | .44 | .71 |
| Animal on Road Query 2 | **.043** | **.033** |
| Animal on Road Query 3 | .34 | .53 |
| Road Closed Query 1 | .74 | .74 |
| Road Closed Query 2 | .22 | > .99 |
| Road Closed Query 3 | .34 | .61 |

*Note. p*-values significant after FDR adjustment are presented in bold.

**Responses to individual Level 3 SAGAT Questions**
**Table 12**
*Significance Tests for the Proportion of Correct Responses for Individual SAGAT Questions for Situation Awareness Level 3*

| Query Name | Haptics (p-values) | Question Presentation Time (p-values) |
|---|---|---|
| Accident Query 1 | .11 | > .99 |
| Accident Query 2 | > .99 | **.012** |
| Accident Query 3 | > .99 | .26 |
| Traffic Jam Query 1 | .69 | .23 |
| Traffic Jam Query 2 | > .99 | > .99 |
| Traffic Jam Query 3 | .05 | .70 |
| Road Construction Query 1 | .69 | > .99 |
| Road Construction Query 2 | .48 | .74 |
| Road Construction Query 3 | .06 | .19 |
| Animal on Road Query 1 | > .99 | .35 |
| Animal on Road Query 2 | > .99 | > .99 |
| Animal on Road Query 3 | .13 | .67 |
| Animal on Road Query 4 | .70 | > .99 |
| Road Closed Query 1 | **.004** | .16 |
| Road Closed Query 2 | .26 | .05 |

*Note. p*-values significant after FDR adjustment are presented in bold.